\documentclass[pre,superscriptaddress]{revtex4}

\usepackage{graphicx}
\usepackage{amssymb}
\def\smallwhitecircle{\mbox{\large $\circ$}}
\def\smallblackcircle{\mbox{\large $\bullet$}}
\def\smallwhitesquare{\mbox{\scriptsize $\square$}}
\def\smallblacksquare{\mbox{\scriptsize $\blacksquare$}}
\def\smallwhitetriangleup{\mbox{\small $\vartriangle$}}
\def\smallblacktriangleup{\mbox{\small $\blacktriangle$}}

\def\bfk{{\bf k}}

\def\xhat{\hat{\bf x}} 
\def\yhat{\hat{\bf y}}

\def\3{\ss }           

\def\bfk{{\bf k}}

\def\bxi{\mbox{\boldmath $\xi$}}
\def\bsxi{\mbox{\boldmath \tiny $\xi$}}
\def\bsigma{\mbox{\boldmath $\sigma$}}

\begin{document}

\title{\large \bf 
Static and dynamic properties of a particle-based algorithm for non-ideal fluids and
binary mixtures}
 
\author{Thomas Ihle$^1$ and Erkan T{\"u}zel$^{2,3}$ 
\\[0.2cm]
$^1$ Department of Physics, North Dakota State University,\\ P.O. Box 5566,
Fargo, ND 58102, USA.\\
$^2$ School of Physics and Astronomy, 116 Church Street SE,\\ University of Minnesota, Minneapolis, MN 55455, USA. \\
$^3$Supercomputing Institute, 
University of Minnesota, \\ 
599 Walter Library, 
117 Pleasant St. SE,
Minneapolis, MN 55455, USA \\
}

\maketitle
 
\vspace{0.1cm}
 
\noindent
A recently introduced particle-based model for fluid dynamics
with effective excluded volume interactions is analyzed in detail.
The interactions are modeled by means of
stochastic multiparticle collisions which are biased and depend on local velocities and densities.
Momentum and energy are exactly conserved locally.
The isotropy and relaxation to equilibrium are analyzed and measured.
It is shown how a discrete-time projection operator technique can be used to obtain Green-Kubo relations for 
the
transport coefficients. 
Because of a large viscosity
no long-time tails 
in the velocity auto-correlation and stress correlation functions were seen.
Strongly reduced self-diffusion due to caging and an order/disorder transition is found at 
high collision frequency, where clouds consisting of at least four particles form a 
cubic phase.
These structures were analyzed by measuring the pair-correlation function above and below the transition.
Finally, the algorithm is extended to binary mixtures which phase-separate above a critical collision rate.

PACS number(s): 47.11.+j, 05.40.+j, 02.70.Ns 

\newpage 

\section{Introduction}

The efficient modeling of the hydrodynamics of complex liquids such as colloidal suspensions
and microemulsions is a challenging task due to the interplay between flow and thermodynamic 
properties and the importance of thermal noise.
The desire to simplify the description of the solvent degrees of freedom led to the development of 
mesoscale simulation methods, which are essentially novel ways of solving the equations of fluctuating hydrodynamics.

In particular, a recently introduced simulation technique of this type --- often called Stochastic 
Rotation Dynamics method (SRD) \cite{ihle_01} or Multi-Particle Collision Dynamics \cite{marisoll_04}--- is a very promising tool for such a coarse-grained description of 
a fluctuating fluid \cite{malev_99,ihle_04}.
The method is based on so-called fluid particles with continuous positions and velocities, which 
follow simple dynamic rules of streaming and collision.
In this sense, SRD is closer to the microscopic reality than conventional Navier-Stokes solvers.
On the one hand, this gives rise to unlimited numerical stability, on the other hand there are certain restrictions on the transport coefficients
which depend on the details of the dynamics.
For instance, the viscosity cannot be chosen to be arbitrarily small or large at reasonable computational efficiency.
Due to the simplicity of the SRD-rules, many analytical calculations could be performed \cite{kikuchi_03,ihle_03} which are very hard to do 
for other particle-based methods.
So far, SRD has been used to study sedimentation \cite{hecht_05,padding_04}, colloids and polymers in flow 
\cite{inoue_02,malev_00}, 
vesicle dynamics 
\cite{noguchi_04}, reacting flow \cite{kapral_04}, and microemulsions \cite{sakai_02}.

The fluid particles of the original SRD method have an ideal gas equation of state. Hence, they are very compressible and the speed of sound, $c_s$,
is low. In order to have negliglible compressibility effects as in real liquids the Mach number has to be kept small, 
which means that there are limits for the flow speed in the simulation.
Recently, we showed how SRD can be modified to achieve a larger speed of sound and a larger viscosity
\cite{ihle_06a,tuzel_06} which makes
simulations of dense gases and liquids more efficient and realistic. 
This was done in a thermodynamically consistent way by introducing generalized excluded volume interactions among
the fluid particles. 
That means, the kinetic energy is still locally conserved and no complications such as temperature drifts are encountered.
This was a first step towards a consistent SRD-model for more general 
liquids with additional attractive interactions and
the possibility of a liquid-gas phase transition.

While (to our knowledge) this has been the first SRD model with a non-ideal equation of state,
there have been several attempts to generalize other particle methods to non-ideal fluids.
The first one is described by Alexander {\it et al}, \cite{garcia_95}, and is based on a modification of the original Direct Simulation Monte Carlo (DSMC) or Bird's algorithm \cite{bird_94}.
It is called the Consistent Boltzmann algorithm (CBA) \cite{garcia_95, garcia_02}.
The authors labeled the original DSMC algorithm as ``inconsistent'' in the sense that it has the transport properties of a hard-sphere gas, 
but obeys an ideal gas equation of state; CBA has the correct hard-sphere equation of state. 
However, back-scattering events connected with structural effects are absent in the model \cite{garcia_95},
leading to too high self-diffusion at higher densities. Furthermore, the sound speed determined from the equal-time density fluctuations are not in agreement with the value obtained directly from the equation of state.
As seen from Fig. 2 and 3 in \cite{garcia_95}, the analytical expressions from Enskog-theory for the transport coefficients are significantly lower 
than the simulation results.

Another particle method very close to DSMC explicitly aimed to solve the 
Enskog equation  is given in Refs.
\cite{nabu_86, montanero_96}.
The method correctly reproduced the transport properties of the Enskog gas, but has the unpleasant feature of conserving momentum and energy only in a statistical sense and not in a single collision.
To overcome these deficiencies a modification of the DSMC method was proposed to solve 
the Enskog-equation by Frezzotti \cite{frezzotti_97}.
The main idea is the same as in \cite{montanero_96}: due to the finite size of a particle, binary collisons do not just take place between particles of the same cell, but also particles from adjacent cells can participate, but now 
energy and momentum are conserved in every collision.

The original SRD method has a similiar issue as DSMC: 
The transport properties can be tuned to that of
an interacting gas, while the equation of state remains ideal.
Hence, a generalization which leads to a tunable equation of state is highly desirable.
This also allows us to observe phase transitions and critical phenomena, 
and the method can be easily extended to multi-component systems.
Furthermore, it was shown that the generalized SRD is ``more consistent'' than CBA 
since independent measurements of the density fluctuations, 
the pressure and the speed of sound are consistent with each other and agree with theory \cite{ihle_06a}.

In CBA, excluded volume effects are modeled by 
introducing an additional displacement of the particles
away from each other 
after every binary collision.
This displacement is in addition to the free streaming displacement, $\tau\,v_i$,
given by the time step $\tau$ and the
actual velocity $v_i$ of particle $i$.
The idea behind this is 
that particles of finite size collide earlier than point-like particles, and hence
would be further apart after time $\tau$.
It is likely that this enhanced particle relocation leads to an additional 
diffusive term in
the continuity equation for the density, which could be the reason for 
thermodynamically inconsistent density fluctuations.
As discussed in \cite{tuzel_06} one has to be careful with respect to
thermodynamic consistency;
often it is not obvious why a certain algorithm fails to produce the correct thermal 
fluctuations.
Therefore, in generalizing SRD
we followed a different strategy than Alexander {\it et al.} \cite{garcia_95}: 
Our main idea is to coarse-grain the collisions between hard-spheres.
This leads to 
additional effects such as
caging and crystallization, and thermodynamically consistent fluctuations can be obtained. 

Instead of costly binary collisions, which is the core of any DSMC-code,
imagine two clouds of particles containing $M_1$ and $M_2$ particles, and having center of mass velocities $u_1$, $u_2$
respectively. The volume of the clouds is assumed to be equal and constant. 
On average, the momentum transfer between these clouds will be high if they are very dense and move at high speed into each other. 
Hence, the average interaction between these clouds depends on the sign and size of $\Delta u\equiv {\bf u_1}-{\bf u_2}$, and on the product of the particle numbers $M_1 M_2$.
In our model, clouds are defined by the particles in a particular cell of a cubic grid.
Two adjacent cells; which we will call a double-cell,will be picked at random and all the particles will undergo a collective collision
which conserves energy and momentum in the double cell with a collision probability proportional to $\theta(\Delta u) f(\Delta u, M_1M_2)$.
It turns out that the equation of state can be tuned by the choice of the function $f$. 

Keeping the underlying lattice as a simple means to determine collision partners allows us to utilize the projection operator 
formalism worked out for original SRD.
This formalism was used to obtain transport coefficients via Green-Kubo relations \cite{ihle_03}.
Very accurate analytical expressions were found even for low particle density and at
low temperatures \cite{ihle_04, tuzel_03} where the mean-free path
is much smaller than the cell size. Both regimes are usually avoided in DSMC, 
mainly because no good 
agreement between Enskog-theory and numerics can be obtained.

\section{Model}

\begin{figure}
\begin{center}
\vspace{2cm}
\includegraphics[width=3.5in,angle=0]{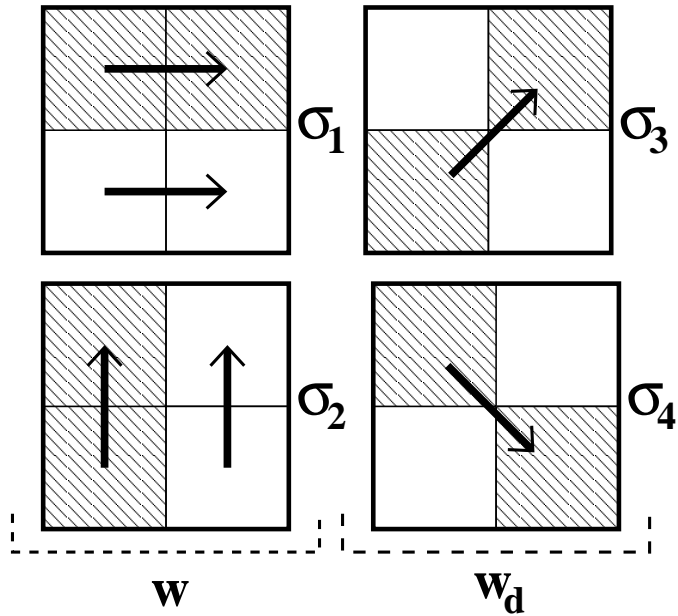}

\caption{Collision rules. Three distinct collisions are considered: (a) horizontally along $\bsigma_1$,
(b) vertically along $\bsigma_2$, (c) diagonally and off-diagonally along $\bsigma_3$ and $\bsigma_4$.
$w$ and $w_d$ are the probabilities of choosing collisions (a), (b), and (c), respectively.}
\label{COLLISION}
\end{center}
\end{figure}

Consider a set on $N$ point-particles with continuous positions ${\bf r}_i$ and velocities ${\bf v}_i$, $i=1...N$.
The particle mass is set equal to one.
The discrete dynamics with time step $\tau$ consists of a streaming and a collision step.
In the streaming step, particles are advected freely with their corresponding velocities, $
{\bf r}_i(t+\tau)={\bf r}_i(t)+\tau\,{\bf v}_i(t)$.
As in original SRD, \cite{malev_99,ihle_01} a cubic grid with lattice constant $a$ is imposed on the system, which is used to organize
multi-particle collisions.
In the current model, another grid with twice the mesh size of the original grid is introduced
which groups four original cells into one supercell (see Fig. \ref{COLLISION}).
For simplicity, we restricted ourselves to two dimensions, but everything can be easily extended to more dimensions.
Each supercell is indexed by the index of the small cell in its lower left corner. 
The origin of the grid is chosen such that the indices of the supercells are always odd.
The position of a cell relative to its supercell is now uniquely determined by whether the components of its index vector $(\xi_x,\xi_y)$
are even or odd. For instance, the cell in the lower right has an even x-index, $\xi_x$, but an odd y-index, $\xi_y$. 

As proposed in \cite{ihle_01} in order to avoid anomalies,
all particles are shifted by the {\it same}
random vector with components in the interval $[-a,a]$ before the collision step
(note this is a larger interval than in original SRD because of the bigger size of the supercells).
Particles are then shifted back by the same amount after the collision.

To initiate a collision, two cells in every supercell are randomly selected.
These two cells form what we will call a {\it double cell}.
The probabilities for the possible double cells have to be determined in a way which is as efficient as possible, i.e. leads to a large
non-ideal part in the equation of state, and furthermore should lead to a model which is isotropic at least at the Navier-Stokes level. 
As will be discussed later, the cubic anisotropy of the underlying grid will show up at the Burnett and higher levels, eventually leading to a cubic
phase at very high collision frequency.

For efficiency, we always pick two double cells in every supercell as possible candidates for momentum exchange.
This way, every particle in the entire system has a chance to be included in a collision. 
As pictured in Fig. \ref{COLLISION} three distinct choices are made in every supercell:
a) horizontal collisions: The two top cells with equal $y$-coordinate form one double cell, and the two lower cells form another double cell.
b) vertical collisions: The two left cells with equal $x$-coordinate form one double cell, and the two right cells form another double cell.
c) diagonal and off-diagonal collisions: The cell from the upper left and the one from the lower right form a double cell, while the remaining two cells 
form the second double cell.

Note, that the diagonal collisions are essential to obtain an equilibration between the kinetic energies in the $x$ and $y$-direction.
Because of $x$-$y$ symmetry, the probabilities for choice a) and b) must be equal, and will be denoted by $w$.
The probability for choice c) is given by $w_d=1-2w$.
The ratio $w/w_d$ has to be determined such that the model becomes as isotropic as possible.
One way to fix this ratio is based on the relaxation of the velocity distribution, and will be discussed in the next section.

\subsection{Collision rules}

Fig. \ref{COLLISION} shows the three choices for double cells.
We define unit vectors $\bsigma_j$, $j=1..4$, which connect the center of the selected cells.
For the two possible double cells aligned with the 
$\xhat$-direction one has $\bsigma_1=\xhat$ and for the vertical choices one has $\bsigma_2=\yhat$.
The diagonal and off-diagonal choices are described by $\bsigma_3=(\xhat+\yhat)/\sqrt{2}$ and 
$\bsigma_4=(\xhat-\yhat)/\sqrt{2}$, respectively. 
In every cell the mean particle velocity is defined by
\begin{equation}
{\bf u}_n={1\over M_n}\,\sum_{i=1}^{M_n}\,{\bf v}_i
\end{equation}
where the sum runs over all particles contained in a given cell, $n$ is the cell index.

Now, the projection of the difference of the mean velocities in the selected cells on $\bsigma_j$ is calculated,
$\Delta u={\bf \bsigma_j}\cdot ({\bf u}_1-{\bf u}_2)$.
If $\Delta u<0$, no collision will be performed. For positive $\Delta u$ a collision will occur
with an acceptance probability which depends on $\Delta u$ and the actual particle numbers in the two cells $M_1$ and $M_2$.
In order to enable analytical calculations and for first tests, the following total acceptance probability was chosen:
\begin{eqnarray}
\label{NONID0}
p_A&=&\Theta(\Delta u)\,\,{\rm tanh}(\Lambda) \,\;\;{\rm with}\;\; 
\Lambda=A\,\Delta u\,M_1M_2
\end{eqnarray}
where $A$ is some parameter to tune the equation of state.
In the limit of $A\rightarrow \infty$ this gives $p_A=\Theta(\Delta u)$, which maximizes the collision frequency, 
leads to a large non-ideal part of the pressure, and thus to a large speed of sound (the more collisions the faster sound can travel).
Unfortunately, as will be shown in a later section, in this limit, the pressure has a non-analytic dependence on density and temperature, which leads to certain thermodynamic inconsistencies.
More details about this limit can be found in Ref. \cite{tuzel_06}.
Another, computationally simpler choice than Eq. (\ref{NONID0})  would be the first term of an expansion of the hyperbolic tangent
for small A: 
$p_A={\rm min}(1, \Theta(\Delta u) \Lambda)$, but the cusp at $p_A=1$ makes accurate analytical calculations of the 
equation of state difficult. 

This procedure represents a coarse-grained version of the collisions in a real gas.
There, mostly binary collisions occur, and they are only possible if the particles approach each other, i.e. if 
$\Delta \tilde{u}=\tilde{\bsigma}_{12}\cdot ({\bf v}_1-{\bf v}_2) >0$. ${\bf v}_i$ is the velocity of the particles, $\bsigma_{12}$ is
the vector connecting their center of mass.
In SRD the generalization of this can be seen as if two clouds of particles coming from the two cells are colliding.
The total momentum transfer of this scattering should be much larger if the two clouds approach each other on average, 
i.e. at $\Delta u>0$, compared to $\Delta u<0$.  
Furthermore, the effective cross section of the scattering should increase with the particle density in the clouds.
This is described by the dependency of the function $\Lambda$ on the particle numbers.
In this model there is large flexibility about how to choose this function with the restriction
that it should be symmetric 
against the interchanging of the two cells.

Once it is decided to perform a collision, the explicit form of the momentum transfer between the two cells is needed.
In close analogy to the hard-sphere liquid, the collision should keep the total momentum and total kinetic energy of the 
double cell invariant, and it should mainly transfer the component of the momentum, which is parallel to the connecting vector
$\bsigma_j$. In the following, this component will be called the parallel or longitudinal momentum, as opposed to the
perpendicular or transverse 
momentum. 
There are many different rules which fulfill these conditions. For instance, a stochastic rotation of the relative 
velocities of 
all particles in the double cells similiar to the rotation rules in SRD for ideal gases, could be used.
Our goal here is to obtain a large speed of sound. 
Therefore, we use a collision rule which leads to the maximum transfer of the parallel momentum, and does 
not change the transverse momentum.
The rule is quite simple, it exchanges the parallel component of the mean velocities of the two cells (parallel to $\bsigma_j$), which is equivalent
to a ``reflection'' of the relative velocities: 
\begin{equation}
\label{NONID2}
v_i^{||}(t+\tau)-u^{||}=-(v_i^{||}(t)-u^{||})
\end{equation}
$u^{||}$ is the parallel component of the mean velocity of the particles of {\it both} cells in a double cell.
Written in Cartesian coordinates this amounts to:

\noindent
a) Horizontal double cells (characterized by $\bsigma_1$):
\begin{eqnarray}
\nonumber
v_{ix}(t+\tau)&=&2u_x-v_{ix}(t) \\
\label{NONID2_1}
v_{iy}(t+\tau)&=&v_{iy}(t)
\end{eqnarray}
b) Vertical double cells ($\bsigma_2$):
\begin{eqnarray}
\nonumber
v_{ix}(t+\tau)&=&v_{ix}(t) \\
v_{iy}(t+\tau)&=&2u_y-v_{iy}(t) 
\end{eqnarray}
c) Diagonal double cell ($\bsigma_3$):
\begin{eqnarray}
\nonumber
v_{ix}(t+\tau)&=&u_x+u_y-v_{iy}(t) \\
v_{iy}(t+\tau)&=&u_x+u_y-v_{ix}(t) 
\end{eqnarray}
d) Off-diagonal double cell ($\bsigma_4$):
\begin{eqnarray}
\nonumber
v_{ix}(t+\tau)&=&u_x-u_y+v_{iy}(t) \\
v_{iy}(t+\tau)&=&u_y-u_x+v_{ix}(t) 
\end{eqnarray}
where ${\bf u}=(M_1{\bf u}_1+M_2{\bf u}_2)/(M_1+M_2)$ is the mean velocity of the double cell, ${\bf u}_1$, ${\bf u}_2$
are the mean velocities of the cells forming the double cell.
One easily verifies that these rules conserve momentum and energy in the double cell.

\subsection{Collision probabilities and isotropy}

The ratio $w/w_d$ describes how often collision cells in the vertical and horizontal direction, a) and b)  
are chosen 
compared to diagonal cell pairs, case c) in Fig. \ref{COLLISION}.
We determined this ratio by requiring that the relaxation of the velocity distribution functions of the particles
 is isotropic and does not depend on the orientation of the underlying grid.

\begin{figure}
\begin{center}
\vspace{2cm}
\includegraphics[width=5in,angle=0]{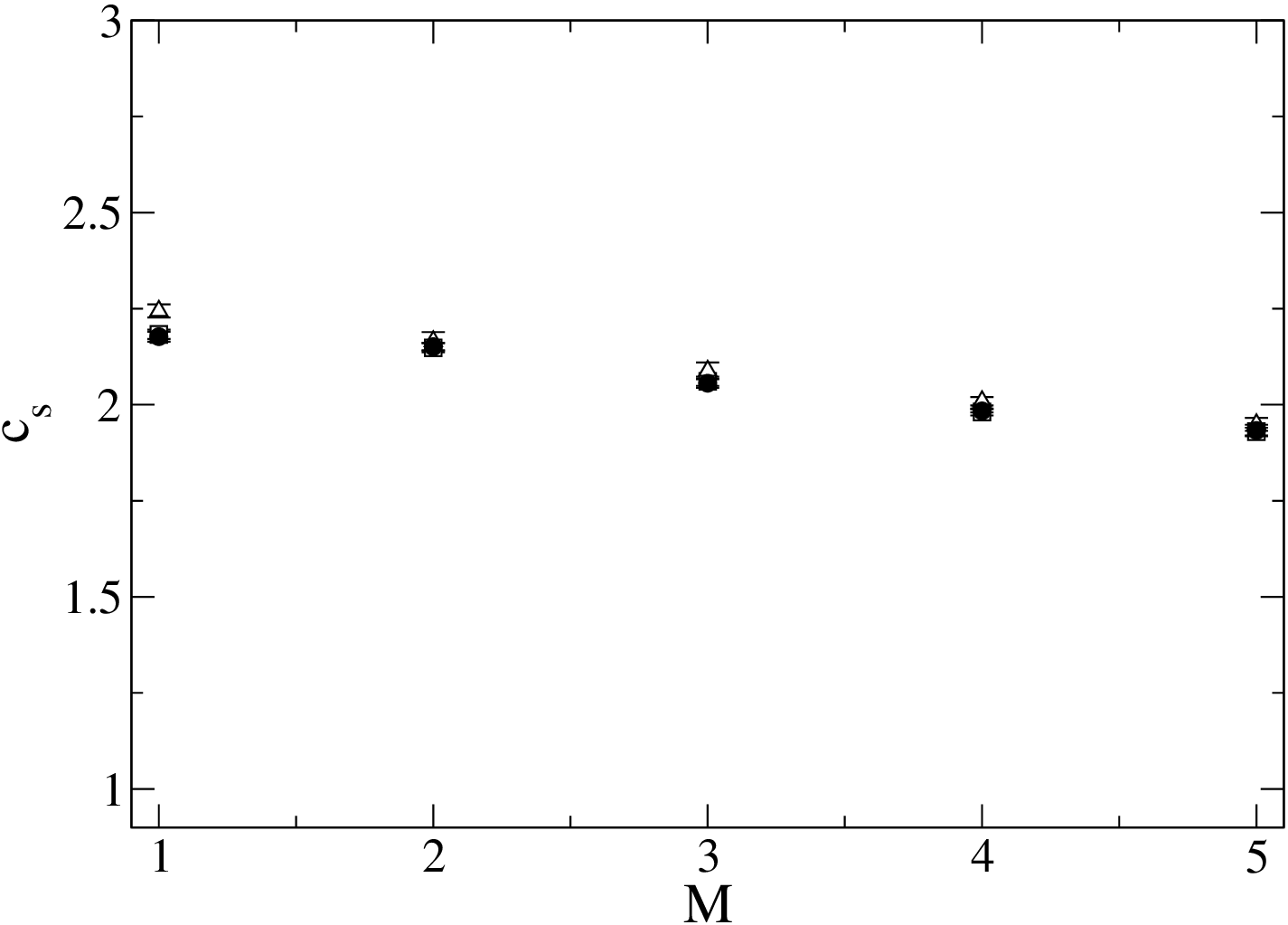}

\caption{Adiabatic speed of sound $c_s$ measured using the time dependent density-density correlations for $\lambda/2a=0.05$.
Densities are obtained at the supercell level and the period of oscillations in the density correlations are used to calculate the speed of sound. Data is shown for the wave vectors $\bfk = (2\pi/L)(1,0)$ ($\smallblackcircle$), $\bfk = (2\pi/L)(0,1)$ ($\smallwhitesquare$) and $\bfk = (2\pi/L)(1,1)$ ($\smallwhitetriangleup$), respectively.
Parameters: $L/a=128$, $A\rightarrow \infty$ and $k_BT=1.0$.}
\label{FIG_SOUND}
\end{center}
\end{figure}

Instead of analyzing the temporal evolution of the entire distribution function, we will restrict ourselves to its lowest moments. 
Assuming molecular chaos for the moment, i.e. that velocities of different particles are uncorrelated at equal times, it is sufficient to 
consider the following three moments of a single particle $i$: $\langle v_{ix}^2\rangle$,
$\langle v_{iy}^2\rangle$, and $\langle v_{ix} v_{iy} \rangle$. For simplicity we will omit the particle label $i$ from now on.
We assume a homogeneous, but non-equilibrium inital state and study the temporal evolution of these moments to their corresponding values given by the Maxwell-Boltzmann
distribution.
We define the vector of second moments as follows:
\begin{equation}
\Psi(t)=
\left( \begin{array}{c}
\langle v_x^2(t) \rangle \\
\langle v_y^2(t) \rangle \\
\langle v_x(t)\,v_y(t)\rangle 
\end{array}
\right)\;\;.
\end{equation}
For simplicity only the limit $A\rightarrow\infty$ in Eq. (\ref{NONID0}) will be discussed in the following.
This means, on average every second collision
attempt is accepted (since the probability to have a positive velocity difference $\Delta u$ is $1/2$).
Using the collision rules, Eq. (\ref{NONID2}), we obtain the following relaxation in one time step:
\begin{equation}
\Psi(t+\tau)=R\,\Psi(t)
\end{equation}
 with the relaxation matrix $R$,
\begin{equation}
R=\left( \begin{array}{ccc} 
2w+(w_d/2)(1+1/n) & (w_d/2)(1-1/n) & 0 \\
(w_d/2)(1-1/n) & 2w+(w_d/2)(1+1/n) & 0 \\
0          & 0        & w_d+2w/n
\end{array}
\right)\;\;.
\end{equation}
Here, $n$ is the actual, fluctuating number of particles in a given {\it double} cell.
Now we consider a rotation of the coordinate system by an arbitrary angle $\beta$, i.e. 
\[ \widehat{\Psi}=\left( \begin{array}{cc} c & s \\ -s & c \end{array} \right)\,\Psi \]

\noindent with $c=\cos{\beta}$ and $s=\sin{\beta}$.
This results in a transformation of the relaxation matrix from $R$ to $\tilde{R}=O\, R\, O^{-1}$ with a matrix 
\begin{equation}
O=\left(
\begin{array}{ccc}
c^2 & s^2 & 2cs \\
s^2 & c^2 & -2cs \\
-cs & cs & c^2-s^2
\end{array}
\right)\;\;.
\end{equation} 
One of the three eigenvalues of $R$ is always one due to the conservation of total energy, i.e. 
$\langle v_x^2\rangle+\langle v_y^2\rangle=\textnormal{const}=2k_B T/m$.
The other two eigenvalues are given by 
\begin{eqnarray}
\nonumber
\lambda_1&=&w_d+{2w\over n} \\
\lambda_2&=&2w+{w_d\over n} 
\label{EIGEN1}
\end{eqnarray}
Requiring isotropy amounts to $\tilde{R}=R$, which is possible only if $\lambda_1=\lambda_2$,
since the rotation of the coordinate system will mix these two modes.
This condition can be fulfilled for arbitrary $n$ only if
$w_d=1/2$ and $w=1/4$. Inserting these values we indeed find that the relaxation matrix is invariant under coordinate rotations: $\tilde{R}=R$. 
Another way to see this is that the requirement of $\tilde{R}=R$ is equivalent to demand that $R$ and $O$ commute:
$[R,O]=RO-OR=0$
This is only possible if $R$ has the following structure with three free parameters $a_1$, $a_2$ and $a_3$:
\begin{equation}
R=\left(
\begin{array}{rrr}
a_1 & a_2 & a_3 \\
a_2 & a_1 & -a_3 \\
-a_3/2 & a_3/2 & a_1-a_2  
\end{array}
\right)\;\;.
\end{equation}
It turns out that original SRD with a rotation angle $\alpha$ is naturally described by a 
relaxation matrix of this shape and that the eigenvalues $\lambda_1$ and $\lambda_2$ are equal.
In the current model, the probabilities $w$ and $w_d$ must be adjusted properly to maintain this isotropic behavior.
Adjusting these probabilities does not neccessarily mean that all properties of the model are isotropic.
This becomes very apparent at high densities or high collision frequency $1/\tau\gg 1$ where one observes inhomogeneuous states with cubic order. 
Other practical tests for isotropy are needed such as
measurements of the speed of sound and of transport coefficients for different directions of
the wave vector ${\bf k}=(k_x, k_y)$.
Fig. \ref{FIG_SOUND} shows the speed of sound measured for three different directions of
the wave vector $\bfk=(2\pi/L)(n_1,n_2)$, with $(n_1,n_2)=(0,1)$, $(1,0)$, and $(1,1)$ where $L$
is the linear dimension of the system. No dependence on the direction was detected, even at this small mean free path $\lambda/(2a)=0.05$, $\lambda=\tau\sqrt{k_B T}$.
A similar measurement (not shown here) revealed that the viscosity for a finite wave vector does not depend on   
the direction of $\bfk$ either.

Choosing $w=1/4$ and $w_d=1/2$ in Eq. (\ref{EIGEN1}) and averaging over the particle number fluctuations
(particles are supposed to be Poisson-distributed as in an ideal gas)
gives the effective decay rate,
\begin{equation}
\label{LAMBDA_EFF}
\lambda_{\textnormal{eff}}=-{1\over \tau}\,{\rm ln}\left({1\over 2}+{1-{\rm e}^{-2M}\over 4M} \right)\,.
\end{equation}
Both, the cross-correlation $\langle v_x (t) v_y(t)\rangle$ and the difference between kinetic energy in x-
and y-direction relax independently with this relaxation rate, i.e.
\begin{eqnarray}
\nonumber
\langle v_x(t) v_y(t)\rangle &=&
\langle v_x(0) v_y(0)\rangle {\rm e}^{-\lambda_{\textnormal{eff}} t} \\
\label{LAMBDA_DEC}
\langle v_x^2(t)\rangle- \langle v_y^2(t)\rangle &=&
\left( \langle v_x^2(0)\rangle- \langle v_y^2(0)\rangle\right) {\rm e}^{-\lambda_{\textnormal{eff}} t}
\end{eqnarray}
\begin{figure}
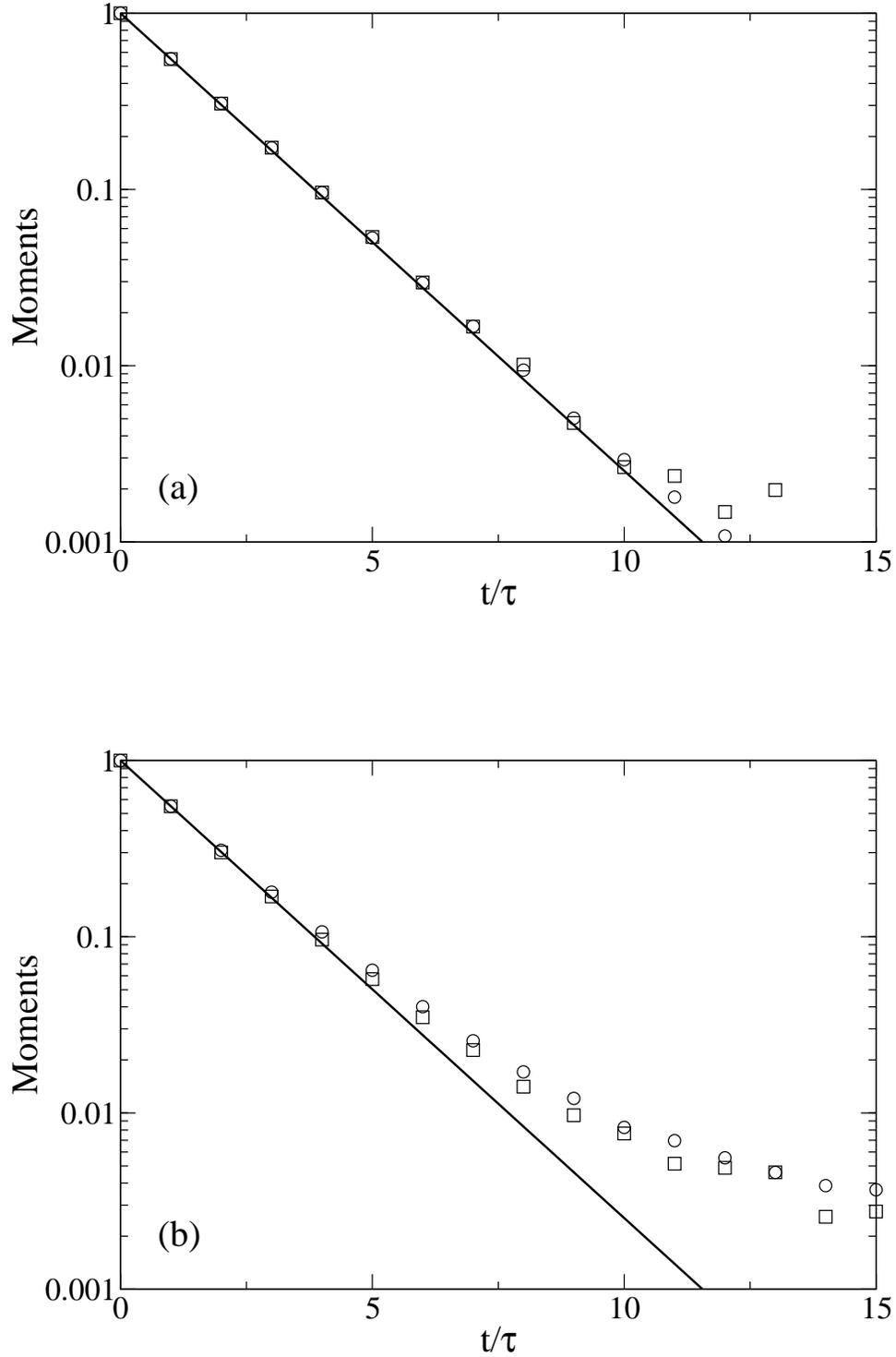

\begin{center}
\vspace{2cm}
\includegraphics[width=5in,angle=0]{moments_vs_t_lambdaover2a_1.50_M5_log.eps}

\vspace{2cm}
\includegraphics[width=5in,angle=0]{moments_vs_t_lambdaover2a_0.05_M5_log.eps}
\caption{Non-equilibrium relaxation of velocity moments
as a function of time.
{\bf (a)} Large mean free path $\lambda/(2a)=1.50$,
{\bf (b)} small mean free path $\lambda/(2a)=0.05$.
Open circles ($\smallwhitecircle$) and squares ($\smallwhitesquare$) show the decay of $\langle v_{x}(t)v_{y}(t)\rangle$ 
and $\langle v_{x}^{2}\rangle -\langle v_{y}^{2}\rangle $ correlations respectively.
The solid lines are the predictions of Eqs. (\ref{LAMBDA_EFF}) and (\ref{LAMBDA_DEC}).
Parameters: $L/a=32$, $M=5$, $A\rightarrow \infty$ and $k_BT=1.0$.}
\label{FIG_MOMENTS}
\end{center}
\end{figure}
\noindent Fig. \ref{FIG_MOMENTS} shows the time relaxation of these moments for small 
and large
mean free path. The initial configuration was set up far from equilibrium
with a large difference between $\langle v_x^2\rangle$ and $\langle v_y^2\rangle$ and
a strong correlation between the x- and y-component of the velocity.
The measured decay is exponential for large mean free path $\lambda/(2a)=1.5$. 
For smaller $\lambda/(2a)=0.05$ one sees a slower decay starting after a few iterations. 
The reason is that the assumption of molecular chaos is invalid at small $\lambda/a$.
This is similar to the behavior of the velocity auto-correlation function 
analyzed in \cite{tuzel_06b}. 

\begin{figure}
\begin{center}
\vspace{2cm}
\includegraphics[width=5in,angle=0]{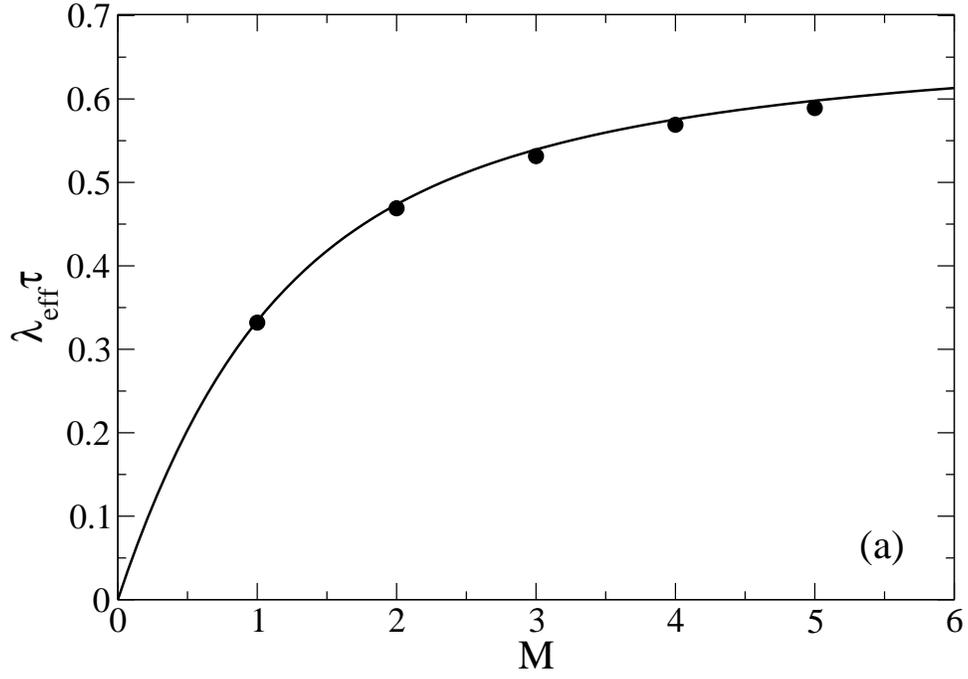}

\vspace{2cm}
\includegraphics[width=5in,angle=0]{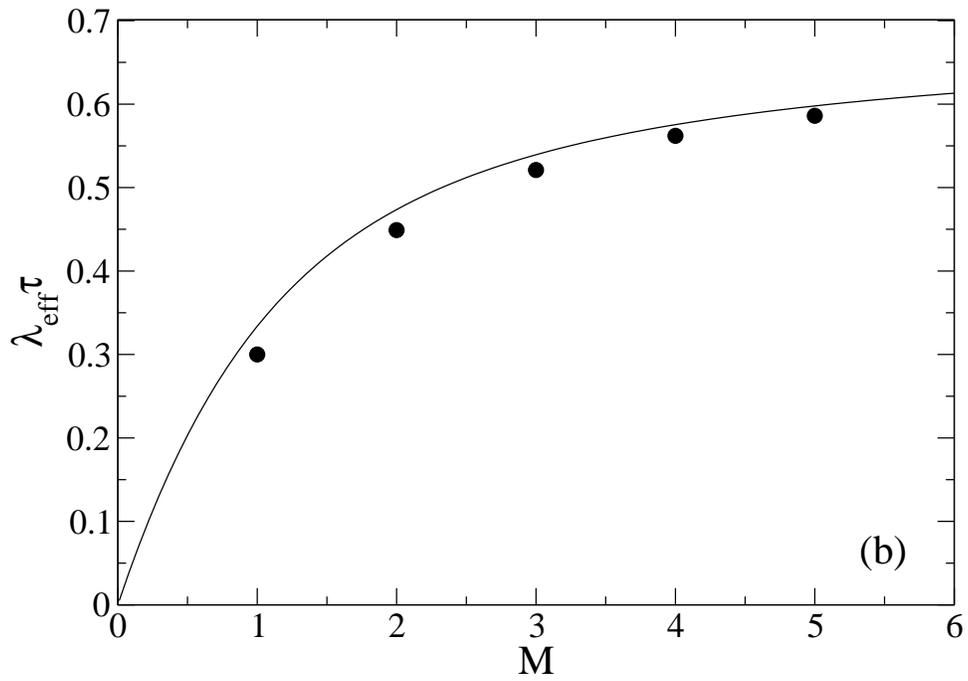}
\caption{Dimensionless relaxation rate $\lambda_{\textnormal{eff}}\,\tau$ for the non-equilibrium 
relaxation of the velocity moments $\langle v_x v_y\rangle$
and $\langle v_x^2\rangle -\langle v_y^2\rangle$ as a function of $M$, the mean particle number per cell.
{\bf (a)} $\lambda/2a=1.50$, {\bf (b)}  $\lambda/2a=0.05$. Bullets ($\smallblackcircle$) show measurements and the solid line is a
plot of Eq. (\ref{LAMBDA_EFF}).
Parameters: $L/a=32$, $M=5$, $A\rightarrow \infty$ and $k_BT=1.0$.}
\label{FIG_EIGEN}
\end{center}
\end{figure}
Fig \ref{FIG_EIGEN} shows the dimensionless decay rate $\lambda_{\textnormal{eff}}\,\tau$ as a function of $M$.
At $\lambda/(2a)=1.5$ very good agreement is found for small and large $M$.
At smaller $\lambda$ the decay is slightly slower due to the mentioned correlation effects.

\subsection{Phase-space contraction and detailed balance}

It is easy to see that the original SRD method conserves phase space and obeys detailed balance \cite{malev_99,ihle_03}.
The situation is more tricky for any coarse-grained, discrete-time method of hard-sphere collisions.
In our model, any state in the 2dN-dimensional phase space is allowed, in contrast to a real hard-sphere system 
where particles cannot overlap.
It is the collision rules described above which make certain states favorable than others.
These rules ought to be biased to obtain a non-ideal equation of state and to 
build up non-ideal correlations between particle positions and their velocities.
On the other hand, kinetic energy is conserved like in a real hard-sphere liquid, and we are in the microcanonical ensemble.
Under these circumstances it is unavoidable that the dynamics must be phase space contracting, it has to project out certain states or at least sample 
them with a smaller probability. These unfavourable states correspond to the ones which are entirely forbidden in a real hard-sphere system.

A superficial analysis
seems to indicate that phase space is conserved: The Jacobian of the streaming step is always one
and the Jacobian of the collision operation is either $+1$ or $-1$.
A closer look reveals, that the dependence of the collision probability on $\Delta u$ destroyes this picture.
Assume a N-particle state $A$ with only one double cell and with $\Delta u<0$. 
No collision is allowed to happen, the state remains unchanged: $A(t+\tau)=A(t)$ (We leave out streaming for the moment).
Consider now a state $B$ with identical particle positions as $A$ but inversed velocities $v_i^B=-v_i^A$.
This time $\Delta u>0$ and a collision happens resulting in $B(t+\tau)=A(t)$.
Hence two different states are mapped onto exactly the same state in one iteration: phase space is contracted.

As a consequence, the dynamics is not time-reversible even in a stochastic sense: Inverting the signs of all velocities of state $B(t+\tau)=A(t+\tau)$ and performing one iteration would lead to a collision going back to the 
time-inverted version of state $B$ even if one would have started at $A$ initially.
In other words, there is no way to go back in time to a state which did not fulfill the collision criteria, namely $\Delta u>0$.

The whole dynamics presumably still fulfills detailed balance or at least semi-detailed balance, which is hard to prove 
since this involves not only the transition probabilities 
between states but also the (difficult to obtain) probability of the equilibrium states itself.
Instead, we checked several thermodynamic properties of the system such as consistency of the thermal fluctuations
with the pressure, \cite{ihle_06a,tuzel_06}, and the speed of sound which are related by thermodynamic expressions, and the scaling of interface fluctuations for the binary version of the model \cite{tuzel_07}.
No inconsistency due to the absence of time-reversibility or due to a possible violation of detailed balance could be observed.

Note, that also neither the consistent Boltzmann algorithm (CBA) \cite{garcia_95}, nor the coarse-grained model by Pooley \cite{pooley_thesis} nor any other method based on the Boltzmann or Enskog equation are time-reversible, for similiar reasons.

\subsection{Additional SRD step}

In order to change the viscosity independently from the equation of state, additional conventional stochastic rotations on the 
cell level can be performed. They will be denoted by the operator $R$. The streaming step with a time step $\tau$ will be denoted by the operator $S(\tau)$ and the biased collisions are called $C$.

To obtain a time-reversal order of operations one has to be careful once the time evolution consists of 
more than 
two non-commuting operations.
Consider the set of operations $...\,S\,C\, R\, S\, C\, R ...$. Inverting time, i.e. reading the sequence from the back leads to the chain
$...S\, R\,C\,S\,R\,C...$ which is not equivalent to the previous one since $C$ comes always after $R$, instead of $R$ always following $C$.
This set of operations can be symmetrized. One possibility would be using streaming with half the time step $\tau/2$ such as
$...S(\tau/2)\, R\, S(\tau/2)\, C\, S(\tau/2) ...$. 
This chain reads the same from the back, since $R$ and $C$ are embedded in $S$-operations.
Another computationally more convenient choice would be random symmetrization:
$...S\,\tilde{R}\, C\, \tilde{R}\, S\, \tilde{R}\, C\, \tilde{R} ...$. 
Here $\tilde{R}$ represents a rotation $R$ which is only chosen with a probability $1/2$.
This is the symmetrization we actually used in all our simulations.
These ways of approximating time evolution operators are very similiar to Trotter-Suzuki formulaes used 
in modern symplectic Molecular Dynamics and Quantum Monte Carlo methods \cite{thijssen_99}.

\section{Equilibrium calculation of transport coefficients}

\subsection{Projection operator formalism}

In Ref. \cite{ihle_03} a projection operator formalism was developed to derive  
the linearized hydrodynamic equations  
from the microscopic collision rules of 
Stochastic Rotation Dynamics (SRD).
This technique was originally introduced by R. Zwanzig 
\cite{zwan_61,mori_65a,mori_65b} and later adapted for lattice-gases by Dufty and Ernst \cite{duft_89}.
With the 
help of this formalism, explicit expressions for both the reversible (Euler)  
as well as dissipative terms of the long-time, large-length-scale hydrodynamics 
equations for the coarse-grained hydrodynamic variables were derived. In addition, 
Green-Kubo (GK) relations were obtained which enable explicit calculations of the transport 
coefficients of the fluid. 
The GK-relations of SRD differ from the well-known continuous versions due to the discrete time-dynamics and the
underlying lattice-structure.

In the following, we will briefly outline how this technique is extended to accomodate 
the new collision rule, Eq. (\ref{NONID2}), which in contrast to SRD does not conserve momentum and energy in {\it single} 
cells but in randomly chosen 
{\it double } cells instead. For more details about the formalism the reader is referred to Ref. \cite{ihle_03}.

The starting point of this theory are microscopic definitions of local, hydrodynamic
variables $A_\beta$. 
These variables are the 
local density, momentum, and energy density.
They can be defined on the cell level as
\begin{eqnarray}
\label{cons}
A_{\beta}(\bxi) &=& 
\sum_{i=1}^{N}
 a_{\beta,i} \prod_{\gamma=1}^d \Theta\left({a\over2}-\vert\xi_\gamma - r_{i\gamma}\vert\right),
\end{eqnarray}        
with the discrete cell coordinates $\bxi=a{\bf m}$,
with $m_\beta=1,\dots,L$, for each spatial component. 
$a_{1,i}=1$ describes the particle density, $\{a_{\beta,i}\}=\{v_{i(\beta-1)}\}$, 
with $\beta=2,...,d+1$, are the components of the particle momenta, and 
$a_{d+2,i}=v_i^2/2$ is the kinetic energy of particle $i$.  
$d$ is the spatial dimension,
${\bf r}_i$ and ${\bf v}_i$ are position and velocity of particle $i$, respectively.
The hydrodynamic variables look much simpler and are easier to treat in Fourier-space.
Their spatial Fourier transforms are given by   
\begin{equation} 
\label{ftc}
A_\beta({\bf k}) = 
\sum_j a_{\beta,j} {\rm e}^{i{\bf k}\cdot\bsxi_j},  
\end{equation} 
where $\bxi_j$ is the coordinate of the cell occupied by particle $j$.  
${\bf k} = 2\pi{\bf n}/(aL)$ is the wave vector, where $n_\beta=0,\pm1,\dots,
\pm(L-1),L$ for the spatial components.  

The next step is to set up an evolution equation in discrete time
which looks like a continuity equation in Fourier-space,
\begin{equation}\label{consl} 
\Delta_tA_\beta(t) + i\bfk\cdot{\bf D}_\beta(t)=0\ ,
\end{equation} 
where $\Delta_t$ is the discrete time evolution operator, defined as $\Delta_t A(t) = [A(t+\tau)-A(t)]/\tau$.
However, up to this point this equation does not contain any information; 
the flux ${\bf D}_\beta$ is 
just formally defined by 
$i\bfk\cdot{\bf D}_\beta(t)=-
[A_\beta(t+\tau)-A_\beta(t)]/\tau$.
The key point in making this expression meaningful
is to set up a microscopic expression for the local
conservation of density, momentum and energy
(for SRD this expression is given by Eq. (23) in Ref. \cite{ihle_03}).
This expression is then inserted into the formal definition of the fluxes ${\bf D}_\beta$ in order to
cancel divergent terms proportional to $1/k$. If this can be done, the reformulated ${\bf D}_\beta$ really 
is a 
flux and Eq. (\ref{consl}) is proven to be a continuity equation for the conserved quantities $A_\beta$.

In the current model, for every choice of double cells,
the local conservation law 
can be expressed as
\begin{equation}
\label{conserv}
\sum_j \left( {\rm e}^{i{\bf k}\cdot{\bsxi}^s_j(t+\tau)}+{\rm e}^{i{\bf k}\cdot({\bsxi}^s_j(t+\tau)+
{\bf z}_{jl})} \right)
[a_{\beta,j}(t+\tau)-
                 a_{\beta,j}(t)]=0,
\end{equation}
where ${\bxi}^s_j$ is the coordinate of the cell occupied by particle $j$
in the {\it shifted} system. Note, that similiar to original SRD, a random shift of cells before collisions
is required to remove anomalies at small mean free path, see Ref. \cite{ihle_01}.
The vector ${\bf z}_{jl}$ is a function of the cell coordinate ${\bxi}^s_j$ and has components which are either $0$, $1$ or $-1$.  
It is constructed such that the sum of the two exponentials in Eq. (\ref{conserv}) is the same for two particles if and only if they are 
in the same double cell. 
The index $l$, $l=1,2...6$ describes the choice of double cells in more detail than the vector $\bsigma_j$:
$l=1$ denotes the lower horizontal double cell, whereas $l=2$ describes the top horizontal double cell
in Fig. \ref{COLLISION}.
Similarly, $l=3$ stands for the left vertical double cell, and $l=4$ is needed if the particle $j$ happens to be in the right vertical double cell.
$l=5$ is for the diagonal choice, and $l=6$ for the off-diagonal choice of double cell.

The components of ${\bf z}_{jl}$ are defined by:
\begin{eqnarray}
\nonumber
z_{j1,x}&=&\left( \begin{array}{rll}
0;       &                             &\xi_{j,y}^s\;\;{\rm  even} \\
1;       &  \xi_{j,x}^s\;\;{\rm  odd}, & \;\xi_{j,y}^s\;\;{\rm odd}                \\
-1;      &  \xi_{j,x}^s\;\;{\rm  even},&\;\xi_{j,y}^s\;\;{\rm odd}                
\end{array} \right. \\
z_{j1,y}&=&0
\end{eqnarray}
\begin{eqnarray}
\nonumber
z_{j2,x}&=&\left( \begin{array}{rll}
0;       &                             &\xi_{j,y}^s\;\;{\rm  odd} \\
1;       &  \xi_{j,x}^s\;\;{\rm  odd}, & \;\xi_{j,y}^s\;\;{\rm even}                \\
-1;      &  \xi_{j,x}^s\;\;{\rm  even},&\;\xi_{j,y}^s\;\;{\rm even}                
\end{array} \right. \\
z_{j2,y}&=&0
\end{eqnarray}

\begin{eqnarray}
\nonumber
z_{j3,x}&=&0 \\
z_{j3,y}&=&\left( \begin{array}{rll}
0;       & \xi_{j,x}^s\;\;{\rm  even},  &                                          \\
1;       &  \xi_{j,x}^s\;\;{\rm  odd}, & \;\xi_{j,y}^s\;\;{\rm odd}                \\
-1;      &  \xi_{j,x}^s\;\;{\rm  odd},&\;\xi_{j,y}^s\;\;{\rm even}                
\end{array} \right. 
\end{eqnarray}
\begin{eqnarray}
\nonumber
z_{j4,x}&=&0 \\
z_{j4,y}&=&\left( \begin{array}{rll}
0;       & \xi_{j,x}^s\;\;{\rm  odd},  &                                          \\
1;       &  \xi_{j,x}^s\;\;{\rm  even}, & \;\xi_{j,y}^s\;\;{\rm odd}                \\
-1;      &  \xi_{j,x}^s\;\;{\rm  even},&\;\xi_{j,y}^s\;\;{\rm even}                
\end{array} \right. 
\end{eqnarray}
\begin{eqnarray}
\nonumber
z_{j5,x}&=&\left( \begin{array}{rll}
0;       & \xi_{j,x}^s\;\;{\rm  even}, & \xi_{j,x}^s\;\;{\rm odd }              \\
0;       & \xi_{j,x}^s\;\;{\rm  odd},  & \xi_{j,x}^s\;\;{\rm even}              \\
1;       & \xi_{j,x}^s\;\;{\rm  odd},  & \xi_{j,x}^s\;\;{\rm odd}              \\
-1;      & \xi_{j,x}^s\;\;{\rm  even}, & \xi_{j,x}^s\;\;{\rm  even}              
\end{array} \right.  \\
z_{j5,y}&=&z_{j5,x} 
\end{eqnarray}
\begin{eqnarray}
\nonumber
z_{j6,x}&=&\left( \begin{array}{rll}
0;       & \xi_{j,x}^s\;\;{\rm  even}, & \xi_{j,x}^s\;\;{\rm even }              \\
0;       & \xi_{j,x}^s\;\;{\rm  odd},  & \xi_{j,x}^s\;\;{\rm odd}              \\
1;       & \xi_{j,x}^s\;\;{\rm  odd},  & \xi_{j,x}^s\;\;{\rm even}              \\
-1;      & \xi_{j,x}^s\;\;{\rm  even}, & \xi_{j,x}^s\;\;{\rm  odd}              
\end{array} \right.  \\
z_{j6,y}&=&-z_{j6,x} 
\end{eqnarray}
In summary, Eq. (\ref{conserv}) is just another way of saying that the mean particle number, the mean velocity and
energy in a double cell is the same before and after a collision.

For the moment we will assume that only one choice for the construction of double cells within a supercell
is possible, i.e. $l$ is kept fixed.
Later, an average over $l$ with the probabilities $w$ and $w_d$ will be performed.
To obtain the evolution equations for the hydrodynamic variables $A_\beta$ in ${\bf k}$-space,
the differential quotient $\Delta_tA_\beta = [A_\beta(t+\tau)-A_\beta(t)]/\tau$ has to be written in a form $\sim i{\bf k}\,{\bf D_\beta}$, where the flux $D_\beta$ is  
of order $O(k^0)$. 
Hence we have
\begin{equation}
\label{defin_flux}
-i\tau {\bf k}\cdot {\bf D}_\beta=A_\beta(t+\tau)-A_\beta(t)= 
\sum_j \left[ \tilde{a}_{\beta,j}{\rm e}^{i{\bf k}\cdot{\bsxi}_j(t+\tau)}
-a_{\beta,j}(t){\rm e}^{i{\bf k}\cdot{\bsxi}_j(t)}
\right]
\end{equation}
with $\tilde{a}_{\beta,j}=a_{\beta,j}(t+\tau)$.
The first term on the r.h.s. can be written as:
\begin{eqnarray}
\nonumber        
\sum_j \tilde{a}_{\beta,j}{\rm e}^{i{\bf k}\cdot{\bsxi}_j(t+\tau)}&=&
\sum_j \tilde{a}_{\beta,j}\left( {\rm e}^{i{\bf k}\cdot{\bsxi}_j(t+\tau)}
+{1\over 2}\left[
{\rm e}^{i{\bf k}\cdot({\bsxi}_j^S(t+\tau)+
{\bf z}_{jl})} 
+{\rm e}^{i{\bf k}\cdot({\bsxi}_j^S(t+\tau)
                       )} \right. \right.           \\
\label{achange} 
&-&\left.\left. {\rm e}^{i{\bf k}\cdot({\bsxi}_j^S(t+\tau)+
{\bf z}_{jl})} 
-{\rm e}^{i{\bf k}\cdot({\bsxi}_j^S(t+\tau)
                       )} 
\right]
\right)
\end{eqnarray}
where we have added and subtracted two identical terms.
A similiar identity can be applied to the second term on the r.h.s. of Eq. (\ref{defin_flux}),
\begin{eqnarray}
\nonumber        
\sum_j a_{\beta,j}(t){\rm e}^{i{\bf k}\cdot{\bsxi}_j(t)}&=&
\sum_j a_{\beta,j}(t)\left( {\rm e}^{i{\bf k}\cdot{\bsxi}_j(t)}
+{1\over 2}\left[
{\rm e}^{i{\bf k}\cdot({\bsxi}_j^S(t+\tau)+
{\bf z}_{jl})} 
+{\rm e}^{i{\bf k}\cdot({\bsxi}_j^S(t+\tau)
                       )} \right. \right.           \\
\label{achange1} 
&-&\left.\left. {\rm e}^{i{\bf k}\cdot({\bsxi}_j^S(t+\tau)+
{\bf z}_{jl})} 
-{\rm e}^{i{\bf k}\cdot({\bsxi}_j^S(t+\tau)
                       )} 
\right]
\right)
\end{eqnarray}
As before, the quantity in the square brackets is identical to zero.
Now, both expressions (\ref{achange}) and (\ref{achange1}) are inserted back into Eq. (\ref{defin_flux}).
Terms coming from the second and third terms in the r.h.s. of Eqs. (\ref{achange}) and (\ref{achange1})
cancel each other because of 
the local conservation law, Eq. (\ref{conserv}).
The remaining terms can be rearranged in the following way,
\begin{eqnarray}
\nonumber
A_\beta(t+\tau)-A_\beta(t)&=& 
\sum_j \, (\tilde{a}_{\beta,j}-a_{\beta,j})\,
\left\{{\rm e}^{i{\bf k}\cdot{\bsxi}_j(t+\tau)}
-{1\over 2}\left[
{\rm e}^{i{\bf k}\cdot({\bsxi}_j^S(t+\tau)+
{\bf z}_{jl})} 
+{\rm e}^{i{\bf k}\cdot({\bsxi}_j^S(t+\tau)
                       )} \right] \right\}           \\
\label{flux_later}
&+&\sum_j \, 
a_{\beta,j}(t)\left(
{\rm e}^{i{\bf k}\cdot{\bsxi}_j(t+\tau)}
-{\rm e}^{i{\bf k}\cdot{\bsxi}_j(t)}
\right)\,.
\end{eqnarray}
This expression is of order ${\bf k}$ for small ${\bf k}$.
Comparison with the definition of the flux ${\bf D}_\beta$ in Eq. (\ref{defin_flux})
and 
expanding the exponentials around ${\bf \xi}_j(t)$ for small ${\bf k}$ shows that the flux ${\bf D_\beta}$ is indeed of
order $O(k^0)$ as expected for hydrodynamic modes,
\begin{eqnarray}
\label{final}
{\bf D}_\beta&=&-{1\over \tau}\sum_j \left\{ 
a_{\beta,j}(t)\Delta {\bxi}_j(t)+\Delta a_{\beta,j}(t)\left[\Delta {\bxi}^s_j(t)-{{\bf z}_{jl} \over 2}\right]+
O(i{\bf k})\right\} {\rm e}^{i{\bf k}\cdot{\bsxi}_j(t)}
\end{eqnarray}
where $\Delta \bxi_j=\bxi_j(t+\tau)-\bxi_j(t)$, 
$\Delta \bxi_j^S=\bxi_j(t+\tau)-\bxi_j^S(t+\tau)$, and $\Delta a_{\beta,j}=a_{\beta,j}(t+\tau)-a_{\beta,j}(t)$. 
This expression turns out to be a simple extension of the flux of the original SRD fluid (Eq. (50) in part I 
\cite{ihle_03}), the only difference being
the vector ${\bf z}_{jl}$ which describes the coupling to a neighbor cell.
Hence, the entire formalism for calculating linearized hydrodynamics and transport coefficients for an ideal SRD-fluid can be applied
without changes except the modified definition of the flux.
In particular, using Eq. (66) of \cite{ihle_03} one easily obtains 
a Green-Kubo relation for the kinematic shear viscosity $\nu$.

Using the resummation procedure described in \cite{ihle_04}, this GK-relation can be rewritten in a form which
is easier to evaluate analytically.
The resummation utilizes the exact Galilean-invariance provided by using random grid shifts and
cancels all terms containing the {\it unshifted} cell positions $\bxi_j$.
Then, the viscosity is obtained in terms of the off-diagonal part of a resummed stress tensor $\sigma_{xy}$, 
\begin{equation}
\label{GK_REL}
\nu={\tau \over N k_B T}{\sum^{\infty}_{n=0}}^{'}
\,\langle \sigma_{xy}(0) \sigma_{xy}(n\tau) \rangle
\end{equation}
The prime means that the first term in the sum is multiplied by a weight $1/2$,
and
\begin{equation}
\label{BCOR1}
\sigma_{xy}(n\,\tau)=\sum_{j=1}^N\,\left\{
v_{jx}(n\tau) v_{jy}(n\tau)+v_{jy}(n\tau) \tilde{B}_{jx}(n\tau)
\right\}
\end{equation}
where
\begin{equation}
\label{BCOR2}
\tilde{B}_{jx}(n\tau)=\xi_{jx}^{S}((n+1)\tau)-\xi_{jx}^{S}(n\tau)-\tau v_{jx}(n\tau)
+{1 \over 2}(z_{jx}((n+1)\tau)-z_{jx}(n\tau))
\end{equation}
Note, that the only formal difference to original SRD \cite{ihle_04} is the appearance of the ${\bf z}$ variables.
Therefore, we write
\begin{equation}
\label{BCOR3}
\tilde{\bf B}(n\tau)={\bf B}(n\tau)+{1\over 2}[{\bf z}((n+1)\tau)-{\bf z}(n\tau)]
\end{equation}
where ${\bf B}(n\tau)$ is the definition used earlier in Ref. \cite{ihle_04}.

An important test for the validity of the new model and the definition of the flux is, whether the reduced 
frequency matrix ${\bf \omega}$ (see Eq. (48) in \cite{ihle_03}) is symmetric.
In \cite{ihle_03} we checked that ${\bf \omega}$ is symmetric 
for the original model.
This matrix contains scalar products of the form $\langle A_\beta|D_\gamma \rangle$, $\langle D_\beta|A_\gamma \rangle$
in the limit of ${\bf k}\rightarrow 0$, i.e.
averages where the flux and hence the vector ${\bf z}_{jl}$ occur only linearly. 
Note, that ${\bf z}_{jl}$ is uncorrelated to the velocity-variables $a_{\beta,j}$. Furthermore, the probability is $1/2$ that 
$\xi_{j\alpha}$ is even or odd.  
Hence,$\langle {\bf z}_{jl} \rangle=0$ for all $l$, and $\langle {\bf z}_{jl} a_{\beta,n} \rangle=0$. 
It turns out that the additional non-ideal term in the flux does not contribute to the reduced
frequency matrix; the matrix is symmetric
and identical to the one of the ideal SRD-fluid.

\subsection{Determining the viscosity}

The Green-Kubo relation given by Eq. (\ref{GK_REL}) was evaluated analytically and numerically. 
If only the first term in the stress tensor, Eq. (\ref{BCOR1}), (which contains terms $\sim v_{jx} v_{jy}$) is used in the evaluation of 
(\ref{GK_REL}) one obtains the so-called {\it kinetic} viscosity $\nu_{kin}$.
Including particle number fluctuations based on the Poisson-distribution, we found
in the limit of large acceptance probability, $A\rightarrow \infty$ (see Eq. (\ref{NONID0})):
\begin{equation}
\label{EQ_KIN_1}
\nu^{\infty}_{kin}={k_B T \tau \over 2}
\left(
{6M+1-{\rm e}^{-2M} \over
2M-1+{\rm e}^{-2M}}
\right)
\end{equation}
$M$ is the average particle number in a cell of size $a$.
For small acceptance probability, i.e. for small $A$, one finds
\begin{equation}
\label{EQ_KIN_2}
\nu_{kin}=k_B T\,\tau\left[
{1\over A M^{3/2}}\sqrt{\pi\over k_B T}-{1\over 2}
\right]
\end{equation}
where we have neglected particle number fluctuations.
Since the first term in Eq. (\ref{EQ_KIN_2}) dominates at typical acceptance rates,
the viscosity
is proportional to the square root of 
temperature as in a real gas.
Interestingly, in this limit it turns out that the self-diffusion constant is exactly equal to the kinetic
viscosity, $D=\nu_{kin}$.
More details about the derivation and numerical comparisons will be published elsewhere.
\begin{figure}
\begin{center}
\vspace{2cm}
\includegraphics[width=5in,angle=0]{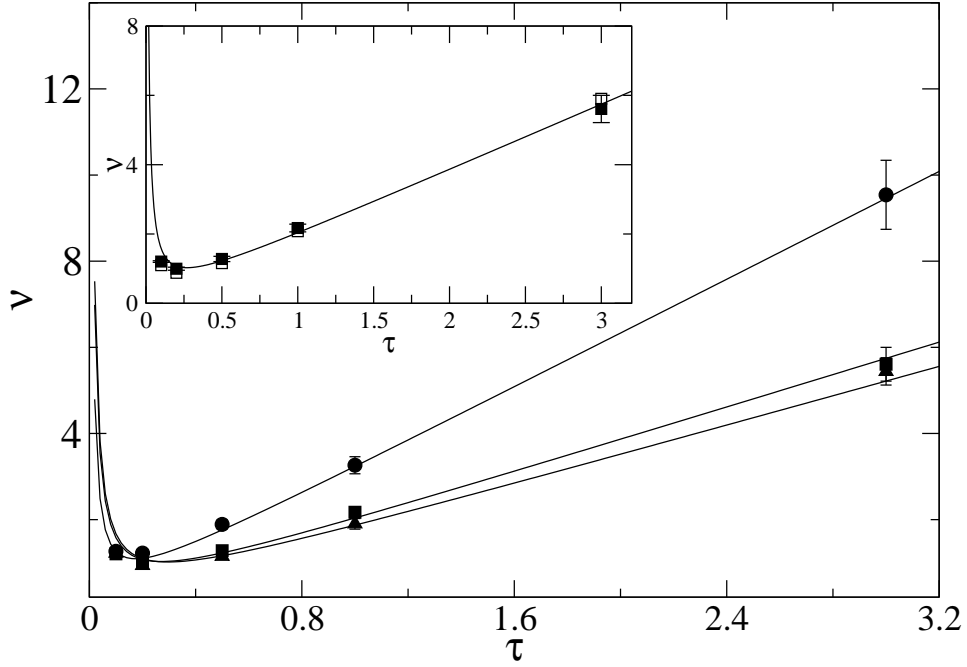}
\caption{ Viscosity as a function of time step $\tau$. Filled circles ($\smallblackcircle$), squares ($\smallblacksquare$) and 
triangles ($\smallblacktriangleup$) show results for $M=1, 3$ and $5$ respectively. Measurements are done fitting vorticity 
decay profiles for the smallest wave vectors and averaging over $5$ different ensembles. The error bars reflect this averaging. 
The inset
shows a comparison of these vorticity measurements ($\smallblacksquare$) with the Green-Kubo measurements ($\smallwhitesquare$) for $M=3$. The solid lines are plots of the sum $\nu_{kin}+\nu_{coll}$ given by Eqs. (\ref{EQ_KIN_1}) and (\ref{EQ_COLL}) in both graphs.
 Parameters: $L/a=32$, $A\rightarrow \infty$ and $k_BT=1.0$.}
\label{FIG_VISCOS}
\end{center}
\end{figure}

For large mean free path $\lambda/(2a)$, the total viscosity is just given by the kinetic viscosity.
However, at small mean free path, additional contributions, denoted by $\nu_{coll}$ become relevant.
For $A\rightarrow \infty$ this contribution takes the form
\begin{equation}
\label{EQ_COLL}
\nu_{coll}=G\,{a^2 \over \tau}\,
\left(1-{1-{\rm e}^{-2M}\over 2M}\right)
\end{equation}
However, the prefactor $G$ is not known yet and might depend
on time step $\tau$ and temperature $T$. Exploring these details is subject of ongoing work.

Figure \ref{FIG_VISCOS} shows measurements of the total viscosity as function of mean free path
and mean particle number per cell.
For comparison, the theoretical expressions for $\nu^{\infty}_{kin}+\nu_{coll}$ from Eq. (\ref{EQ_KIN_1}) and (\ref{EQ_COLL})
are shown with an arbitrary constant prefactor $G=1/6$. Good agreement is found for mean free paths $\lambda/(2a)>0.25$.
The inset in Fig. \ref{FIG_VISCOS} compares measurements of the viscosity using two very different concepts, (i)
using vorticity correlations as explained in Ref. \cite{ihle_01}, and (ii) using the GK-relation (\ref{GK_REL}).
The results of both methods agree within a few percent.

\subsection{Absence of long-time tails}

\begin{figure}
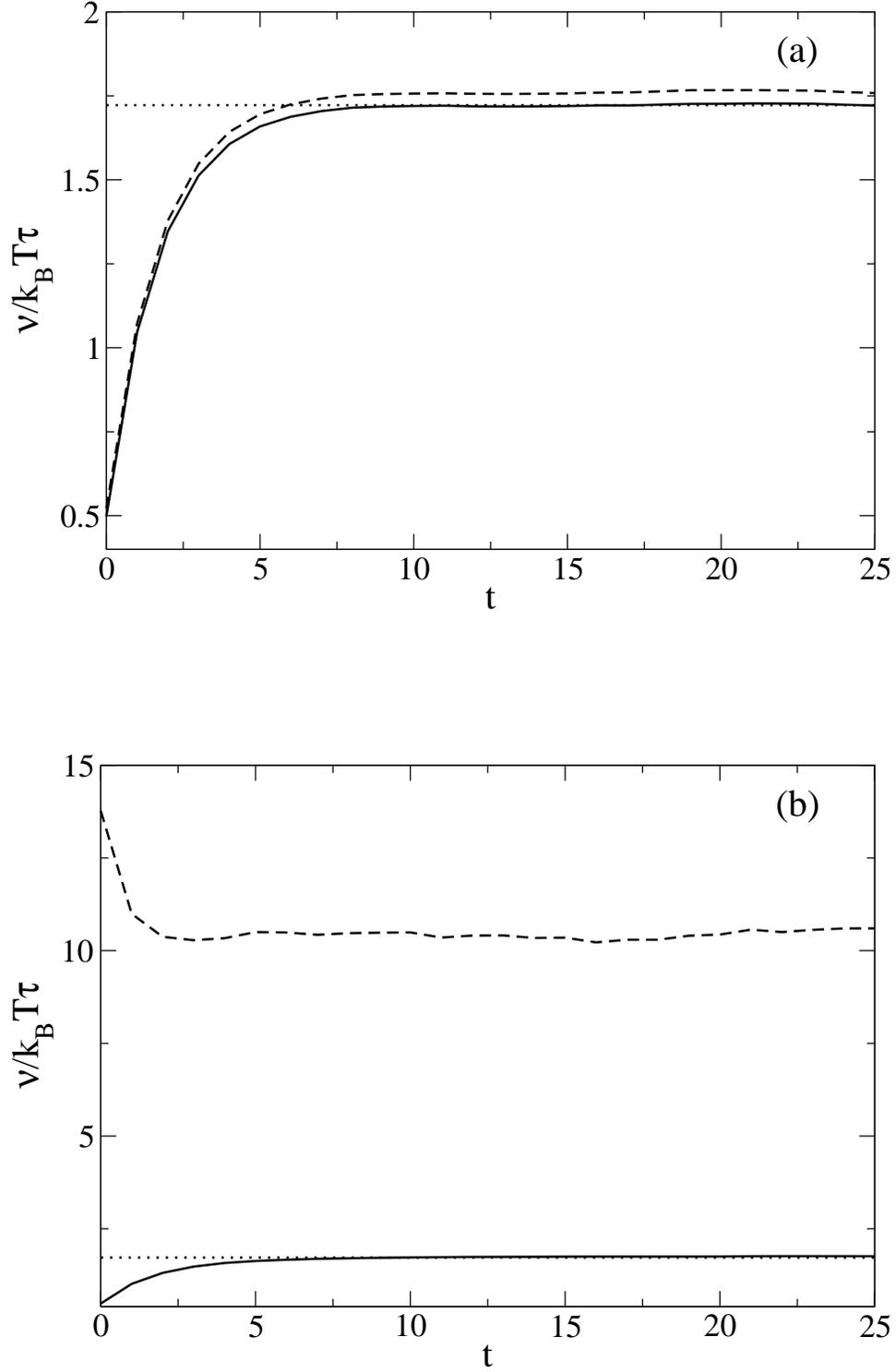

\begin{center}
\vspace{2cm}
\includegraphics[width=5in,angle=0]{Kubo_nuxy_1.eps}

\vspace{2cm}
\includegraphics[width=5in,angle=0]{Kubo_nuxy_5.eps}

\caption{ Temporal evolution of the shear viscosity. 
(a) $\lambda/2a=1.50$, (b)  $\lambda/2a=0.05$. The solid and dashed lines show the kinetic and total contributions to shear viscosity as a function of time step, respectively. The dotted lines show the prediction of Eq. (\ref{EQ_KIN_1}) for $\nu^{\infty}_{kin}$.
Parameters: $L/a=32$, $M=5$, $A\rightarrow \infty$ and $k_BT=1.0$.}
\label{FIG_VISCTAIL}
\end{center}
\end{figure}

In two dimensions, the velocity-autocorrelation function $R(t)=\langle v_{ix}(0) v_{ix}(t) \rangle$,
and the stress correlation function $\langle \sigma_{xy}(0)\sigma_{xy}(t)\rangle$ are expected to decay as $1/t$ at large times.
Since the transport coefficients are related to the integrals of these correlation functions by GK-relations, one should see
a logarithmic divergence of the time-dependent viscosity. 
These ``long-time tails'' and the logarithmic divergence have been measured in the original SRD-method \cite{ihle_03} and found to be in 
quantitative agreement with
general hydrodynamic theory \cite{ernst_70,ernst_71,fors_75}.
However, as can be seen 
in Fig. \ref{FIG_VISCTAIL} no such divergence could be detected in the present model. Instead, the viscosity quickly converges
to a plateau and does not drift with time.
The amplitudes of the expected divergent terms were calculated analytically in the 70's \cite{ernst_70,ernst_71} and shown
to be essentially  
inversely proportional to the density and to the transport coefficients such as the shear viscosity and the self-diffusion constant.
\begin{figure}
\begin{center}
\vspace{2cm}
\includegraphics[width=5in,angle=0]{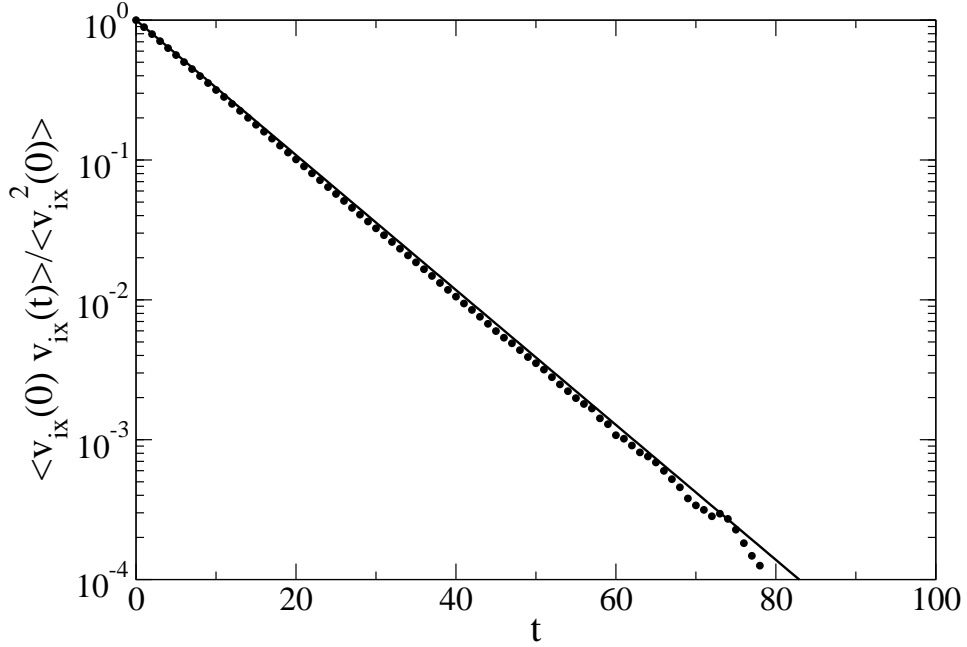}

\caption{Normalized velocity auto-correlation function as a function of time for $\lambda/2a=0.25$. 
The solid line is the prediction of Eq. (\ref{EQ_AUTO}). Parameters: $L/a=32$, $A=1/60$, $M=5$ and $k_BT=1.0$.}
\label{FIG_AUTO}
\end{center}
\end{figure}
We estimated these amplitudes and found that they are too small to be seen within our numerical resolution, 
since the present model is more viscous
than original SRD. 

Fig. \ref{FIG_AUTO} shows the velocity auto-correlation function. For small $A$, a calculation based on the assumption of molecular chaos
gives the following exponential behavior:
\begin{equation}
\label{EQ_AUTO}
R(n\,\tau)=k_B T\left(1-A\sqrt{k_B T \over \pi}M^{3/2} \right)^n
\end{equation}
The measured decay of the auto-correlation function $R(t)$ is exponential over four 
decades and agrees perfectly with expression (\ref{EQ_AUTO}), i.e.
there is no sign of
long-time tails.

\section{The equation of state}

\subsection{The mechanical route to pressure}

In order to obtain explicit hydrodynamic equations, the susceptibility matrix must be known.
These equal time correlations do not follow from the projection operator technique and must be determined by other means.
For a hard-sphere fluid, the only quantity needed to obtain all necessary information is the equation of state.
The other quantity, the internal energy, is already known since it is identical to that of an ideal gas.
This is because we constructed our collision operations to leave the kinetic energy invariant in a double cell.
Here, we follow the mechanical route to pressure, which is defined as the average longitudinal momentum 
transfer
across a fixed interface per unit time and unit surface.

Let us assume this interface to be parallel to the y-direction, and only consider transfer of x-momentum,
i.e.
the component $p_{xx}$ of the pressure tensor is to be determined.
Double cells aligned with the y-direction do not contribute, only double cells characterized
by $\bsigma_1$, $\bsigma_3$, and $\bsigma_4$ have to be considered.
We will only discuss momentum transfer caused by collisions, but not the contributions 
from streaming
which give rise to the ideal part of the pressure.
\begin{figure}
\begin{center}
\vspace{2cm}
\includegraphics[width=5in,angle=0]{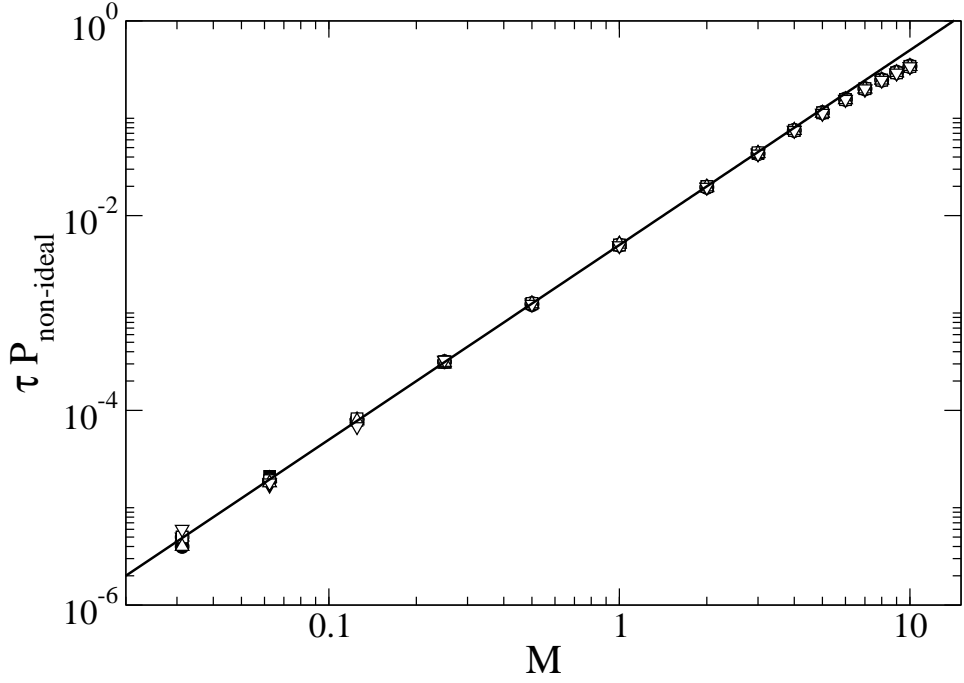}
\caption{Non-ideal part of pressure times $\tau$ as a function of $M$. Data for $\tau=0.05, 0.10, 0.20, 0.40, 0.60,
0.80, 1.00, 2.00$ are collapsed. The solid line is the first term ($\sim k_B T$) of the theoretical prediction of Eq. (\ref{PRESS_MAV}). Parameters: $L/a=32$, $A=1/60$ and $k_BT=1.0$.}
\label{FIG_PRESS}
\end{center}
\end{figure}

Since the collision rules are applied in homogeneously shifted cells, the distance $\delta_x$ between the left most corner of a double cell
and the dividing line in x-direction is  homogeneously distributed between $0$ and $2a$.
The average amount of momentum transferred across the dividing line is zero for $\delta_x=0$ and $\delta_x=2a$, and increases
linearly towards $\delta_x=a$, i.e. reaches a maximum if the dividing line goes through the center of the double cell.
Averaging over this position dependence leads to
$G_{tot}={1\over 2} G_{mid}$. $G_{tot}$ is the averaged transferred momentum, $G_{mid}$ is the transferred
momentum across a line going through
the center of mass of the double cell. 

Let us consider horizontal double cells first.
The change in x-velocity of a particle in one cell during an accepted collision is given by (\ref{NONID2_1}),
\begin{equation}
\label{PRESS1}
\Delta v_{ix}=2\left(u_x-v_{ix}\right),
\end{equation}
hence, the x-momentum change in the left cell is
\begin{equation}
\label{PRESS2}
\Delta G_x=-m{2M_1 M_2 \over M_1+M_2}\Delta u,
\end{equation}
$\Delta u=u_{1x}-u_{2x}$, $u_{1x}$ and $u_{2x}$ being the average x-velocity in the left and right cell, respectively. 
$m$ is the mass of the particles which is set to one.

For the moment we will assume, that the particle numbers $M_1$ and $M_2$ will be constant, i.e. the following  
averages are performed under this restriction.
To obtain the pressure, the thermal average over the momentum transfer across a line is required which involves knowledge of the
acceptance probability $p_A(M_1,M_2,\Delta u)$ for a collision. 
One obtains
\begin{equation}
\label{PRESS3}
\langle \Delta G_x \rangle={w\over 2} \int_0^\infty\,p_G(\Delta u)\,p_A(M_1,M_2,\Delta u) \Delta G_x \,d(\Delta u)
\end{equation}
The factor $1/2$ comes from the position average of the dividing line; the integral involves only positive $\Delta u$, because
the acceptance rate is zero for $\Delta u<0$. $p_G(\Delta u)$ is the probability that $u_{1x}-u_{2x}$ for the micro-state of two
cells is equal to $\Delta u$.
$w$ is the probability to select this double cell and will be equal to $1/4$ as discussed before to ensure isotropy.
One has
\begin{equation}
\label{PRESS4}
p_G(\Delta u)=\int_{-\infty}^\infty\prod_{i=1}^{M_1+M_2}dv_{ix} g(v_{ix})
\delta\left({1\over M_1} \sum_{j=1}^{M_1}v_{jx}-{1\over M_2} \sum_{j=M_1+1}^{M_1+M_2}v_{jx} \right)\,
\end{equation}
where $g(v_{ix})$ is the Boltzmann-weight, $g(v_{ix})\sim{\rm exp}(mv_{ix}^2/2k_B T)$.
Using the integral representation of the $\delta$-function,
\begin{equation}
\label{PRESS5}
\delta(x)=\int_{-\infty}^\infty{\rm e}^{ikx}\,{dk\over 2\pi}
\end{equation}
Eq. (\ref{PRESS4}) leads to 
\begin{equation}
\label{PRESS6}
p_G(\Delta u)=\sqrt{{m\over 2\pi k_B T \gamma}}\,{\rm e}^{{m(\Delta u)^2\over 2\gamma k_B T}},
\end{equation}
with $\gamma=(M_1+M_2)/M_1M_2$.

Expanding the acceptance probability, Eq. (\ref{NONID0}), in $\Lambda=A\,\Delta u\,M_1 M_2$ leads to
\begin{equation}
\label{PRESS7}
p_A(M_1,M_2,\Delta u)=\theta(\Delta u)
\left( \Lambda-{\Lambda^3 \over 3}+...
\right)\,.
\end{equation}
The contributions to the pressure from all terms of this series can be exactly calculated but we restrict ourselves to the first two terms. The reason is that all terms which are non-linear in $\Lambda$ are related to a thermodynamic inconsistency and have to be kept 
small anyways. The size of the first non-linear contribution will give us clues about the useful parameter range and estimations of the violation of thermodynamics. 

Defining the pressure as the average momentum transfer per unit area and unit time,
$p_{1}=\langle \Delta G_x \rangle/(\tau a^{d-1})$ ($d$ is the spatial dimension, which is two in our case), 
the pressure follows from Eqs. (\ref{PRESS3},\ref{PRESS6})
as
\begin{equation}
\label{PRESS8}
p_1={w\, A \over 2}{k_B T\over a\tau}\, M_1 \,M_2
-{w \, A^3 \over 2}{(k_B T)^2\over a\tau} \gamma\,M_1^3 \,M_2^3+...
\end{equation}
The index 1 for the pressure denotes the contribution from horizontal double cells.
A calculation following the same lines yields that  
$p_3$ (due to the diagonal collision $\bsigma_3$) 
can be obtained by just replacing $w$ by $w_d/\sqrt{2}$ in the expression for $p_1$ and by symmetry 
we have $p_3=p_4$.

Using $w=1/4$ and $w_d=1/2$, the total non-ideal pressure under the restriction of the particle numbers in the two cells to be $M_1$ and $M_2$ 
is therefore
\begin{equation}
\label{PRESS9}
P=
\left({1 \over 2 \sqrt{2}}+{1 \over 4} \right) \left[ {A \over 2}{k_B T\over a^{d-1}\tau}\, M_1 \,M_2
-{A^3 \over 2}{(k_B T)^2\over a^{d-1}\tau} (M_1^3 \,M_2^2+M_1^2\,M_2^3)+...\right] \;\;.
\end{equation}
For small particle densities $M_n<4$, the fluctuations of the particle number in a cell cannot be neclected.

As a first approximation, we assume the particles to be Poisson-distributed,
\begin{equation}
\label{POISSON}
p(M_1)={\rm e}^{-M}\,{ M^{M_1} \over M_1!},\;\;\;M_1=0,1,2... 
\end{equation}
$p(M_1)$ is the probability to find $M_1$ particles in a given cell, where $M=\langle M_1\rangle$ is the average number of particles in that cell, and we further assume the distributions of adjacent cells to be independent: $p(M_1, M_2)=p(M_1)\,p(M_2)$.
These assumptions are only strictly true for an ideal gas, but turn out to
be sufficiently accurate. 
In principle, the corrections of this to the non-ideal equation of state and the resulting correlations of the particle numbers in adjacent cells should be calculated self-consistently. 

With these approximations, one obtains the non-ideal part of the pressure averaged over particle number fluctuations
in two dimensions:
\begin{equation}
\label{PRESS_MAV}
P_{non-id}=\left({1 \over 2 \sqrt{2}}+{1 \over 4} \right) {A\,M^2 \over 2} {k_B T\over a \tau}
\left[
1-2A^2M^3\,k_B T\left(1+{4 \over M} + {4 \over M^2} + {1 \over M^3} \right)
+... \right] \;\;.
\end{equation}
\noindent
Note, that $P_{non-id}$ is quadratic in the particle density $\rho=M/a^2$ to lowest order in $\rho$ and for small $A$ as one would expect from a virial expansion.
Furthermore, the first term in the non-ideal pressure is linear in temperature $T$ which is also expected for a hard-sphere gas.
The second term is just the first contribution arising from the non-linearity of the hyperbolic tangent in the acceptance probability. All these terms lead to a non-linear dependency on temperature, which will be shown to be inconsistent with the internal energy
of a hard-sphere gas.
This means, the prefactor $A$ has to be chosen small enough to be in the linear regime. This is equivalent to having a small acceptance rate for a collision. It turns out that acceptance rates of about $15\%$ are sufficiently small to avoid inconsistencies.

In the limit $A\rightarrow \infty$, the acceptance rate is $50\%$ because $\Delta u$ has to be positive
to accept a collision.
In this limit and for large $M$ an expression can be derived which shows a non-analytic dependence of both density and temperature,
namely $P_{non-id}\sim \sqrt{k_B T \rho}$ which again is thermodynamically inconsistent as discussed above.
Another technical difficulty for this case is that the average over the particle fluctuations cannot be performed exactly anymore,
only at large densities an asymptotic result is found. On the other hand, even for arbitrary
values of $A$
the pressure can be obtained as an
infinite sum by expanding the hyperbolic tangent.
All terms in this sum can be calculated exactly.

The off-diagonal terms of the pressure tensor can be calculated along the same lines.
One finds $p_{1xy}=0$, $p_{3xy}=p_{1xx}=-p_{4xy}$.
Hence the total off-diagonal term $P_{xy}$ vanishes.
In the limit of $A\rightarrow \infty$ and for large M, the equation of state has the following non-analytic form:
\begin{equation}
P^{\infty}_{non-id}=\rho k_B T\left(
1+{a \over \lambda} {1\over \sqrt{M}} {1\over \sqrt{\pi}} \left[ {1\over 8}+{1\over 4\sqrt{2}} \right]
\right)
\end{equation}
where $a$ is the lattice unit of the cell, $\lambda$ is the mean free path $\lambda=\tau\sqrt{k_B T}$.

\subsection{A microscopic expression for the pressure}

The microscopic expression for the flux ${\bf D}_\beta$ was derived in Eq. (\ref{final}). The components $\beta=2,..1+d$ also provide a microscopic expression of the stress tensor: the average of the diagonal part gives the virial expression for the pressure, 
\begin{eqnarray}
\nonumber
P_{tot}&=&P_{ideal}+P_{non-id} \\
\label{VIRIAL}
&=&{1\over \tau \, V}
\Big\langle
\sum_j \left\{ 
v_{x,j}\Delta \xi_{x,j}+\Delta v_{x,j}\left[\Delta  \xi^s_{x,j}-{ z_{x,jl} \over 2}\right]
\right\}
\Big\rangle \;\;.
\end{eqnarray}
The first term was discussed in part II of Ref. \cite{ihle_03}, $\langle v_{x,j}\Delta \xi_{x,j} \rangle=\langle \tau v_{x,j}^2 \rangle$
and gives the ideal part of the pressure, $P_{ideal}$. The average over the second term vanishes (see Ref. \cite{ihle_03}); and it is the third term leading to an expression for the non-ideal part of the pressure.
As seen directly from Eq. (\ref{VIRIAL}), this term counts the momentum transfer between the two cells in a double cell:

Consider a horizontal double cell for simplicity. Let $-\Delta p$ be the total momentum change (in x-direction) in the left cell.
Due to momentum conservation in the double cell, the total momentum change in the right cell is equal to $\Delta p$.
Now, $z_x$ is equal to $1$ in the left cell and equal to $-1$ in the right cell.
Hence, as a result of the multiplication by $z_x/2$ in Eq. (\ref{VIRIAL}) we 
have $\Delta p (1/2)-\Delta p (-1/2)=\Delta p$, i.e.
the non-ideal part of the pressure is expressed as the sum of all the momenta which are transferred from the left cell to the 
right cell in all selected double cells in the system. This transferred momentum is mostly positive, since only the collisions with a 
higher mean velocity in the left cell are allowed to happen.
This explains why biased collisions are needed to obtain a non-ideal equation of state; in original SRD there is still
momentum transfer due to rotations across a plane but with positive and negative contributions which cancel each other.

In the current model, due to particle number fluctuations, there can be situations where the mean velocity in the left cell is higher than in the right allowing
a collision, but the mean momentum $=u_1\,M_1$ is smaller than in the right cell with $M_1<M_2$. 
This would lead to a negative contribution to the pressure. 
However, these are events of small probability and are even less relevant at higher particle density. On average, 
unlike in original SRD, the amount of transferred momentum does not vanish, and we have indeed a non-zero pressure.

Comparison with the analytical expression, (\ref{PRESS_MAV}) shows good agreement, Fig. \ref{FIG_PRESS}.
There are deviations at large $M$ because the acceptance rate gets larger with $M$, and at rates above $15\%$
the terms proportional to $(k_BT)^2$ become important in the pressure, (\ref{PRESS_MAV}). These corrections
were not
included in the theory plotted in the figure because the model becomes thermodynamically inconsistent in that range.
Better agreement at larger $M$ can always be obtained by decreasing the prefactor $A$
in Eq. (\ref{NONID0}).

\subsection{Density fluctuations and thermodynamic consistency}

\begin{figure}
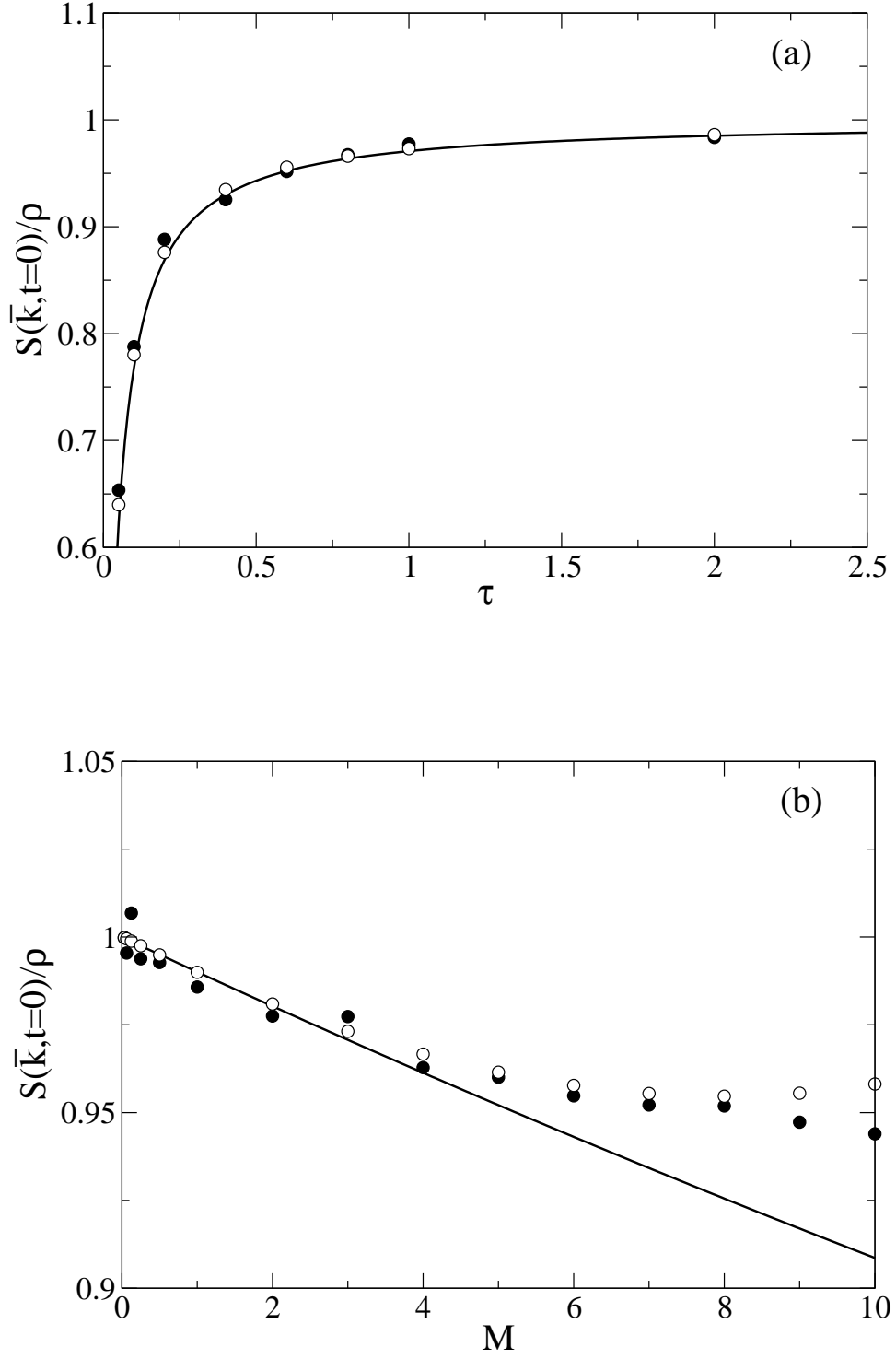

\begin{center}
\vspace{2cm}
\includegraphics[width=5in,angle=0]{rhok_vs_tau_M3.000.eps}
\vspace{2cm}

\includegraphics[width=5in,angle=0]{rhok_vs_M_tau1.00.eps}
\caption{ 
Static structure factor $S(\bar{k},t=0)$ as a function of (a) $\tau$ for $M=3$, (b) $M$ for $\tau=1.0$. 
The open ($\smallwhitecircle$) circles show measurements 
from taking the numerical derivative of the pressure.
The filled ($\smallblackcircle$) circles show direct measurements of the density fluctuations. 
The solid line shows the theoretical prediction based on Eqs. (\ref{PRESS_MAV}) and (\ref{FLUCT1}). 
$\bar{k}$ denotes the lowest possible wave vector, $\bar{k}=(2\pi/L)\,(1,0)$.
Parameters: $L/a=32$, $A=1/60$ and $k_BT=1.0$.}
\label{FIG_RHO1}
\end{center}
\end{figure}

Thermodynamics gives a relation between derivatives of the pressure $p$ and the structure factor $S(k,t)$
at zero time $t=0$ and in the small wave number $|k|\rightarrow 0$ limit,
\begin{equation}
\label{FLUCT1}
S(k,t=0)=\left.\rho k_B T{\partial \rho\over \partial p}\right|_T \;\;.
\end{equation}
On the other hand $S(k,t=0)$ is given by the equal time correlations of the density fluctuations
in Fourier space,
$\langle \rho_k(0) \rho_{-k}(0) \rangle$.
Hence, a numerical consistency check can be performed: The density fluctuations are measured, 
and 
the derivative of the measured pressure is taken numerically and inserted into Eq. (\ref{FLUCT1}).
Both routes should lead to the same $S(k,t=0)$ which can also be calculated analytically using the equation of 
state, (\ref{PRESS_MAV}).
As shown in Fig. \ref{FIG_RHO1}, this works nicely for small prefactors $A$ and both large and small time steps
$\tau$. 
For small $M$ $(<5)$ good agreement is achieved.
For larger $M$, one comes closer to the limit of $A\rightarrow \infty$,
where the acceptance probability is independent of particle densities; 
this limit is thermodynamically inconsistent as we will discuss below.
Of course, if agreement for $M>4$ is required, all one has to do is to reduce $A$, i.e.  
decrease the acceptance probability.

As mentioned before, at large acceptance probability, where 
the non-linear terms in $A$ in the equation of state are not negligible, the model is not thermodynamically consistent.
The reason is that these terms contain non-linear functions of temperature $k_B T$.
This leads to the following problem:
let us assume a term in the equation of state of the form $T^2 \rho^n$, where $n\neq 1$.
The pressure is related to the free energy density $f$ by the relation
\begin{equation}
P(T,\rho)=\rho{\,\partial f\over \partial \rho}|_T-f \;\;.
\end{equation}
This is a linear differential equation for $f$ with the general solution
\begin{equation}
f=C(T)\rho+{T^2\rho^n\over n-1}
\end{equation}
where $C$ is an arbitrary function of temperature.
The entropy density is defined by the following derivative of the free energy density:
\begin{equation}
s=-{\partial f\over \partial T}|_{\rho}
\end{equation}
which gives in our case:
\begin{equation}
s=-{\partial C\over \partial T}\rho-{2T\rho^n\over n-1} \;\;.
\end{equation}
Finally, we obtain the part of the internal energy density related to the pressure term $\sim T^2$ as
\begin{equation}
u=f+sT=\rho\left(C-T{\partial C\over \partial T}\right)-{T^2\rho^n\over n-1}
\end{equation}
The first term is linear in $\rho$, the second one is nonlinear in $\rho$.
Hence it is impossible to have $u=0$ at all densities even for an arbitrary function $C(T)$.
This is a contradiction, because we have a model with collision rules which exactly conserve kinetic energy.
Hence, the internal energy should be the same as for an ideal gas, i.e., there should be a kinetic term only and the additional contribution
related to the non-ideal part of the pressure should be zero at any temperature and density.
The fact that there are additional contributions to the internal energy probably means that the thermodynamic temperature does not agree with the kinetic one (defined via the square of the particle velocities) and might be the cause of the observed inconsistencies in the density fluctuations at large $A$ (defined in Eq. (\ref{NONID0})).
This seems to be supported by another contradiction we observed at large $A$: The measured temperature fluctuations seem to be consistent with the specific heat $c_V=dk_B/2$ of an 
ideal gas, but the theoretical prediction for $c_V$ based on a free energy density with non-linear temperature dependence
differs from $dk_B/2$. 

\begin{figure}[th!]
\begin{center}
\vspace{1.5cm}
\includegraphics[width=5in,angle=0]{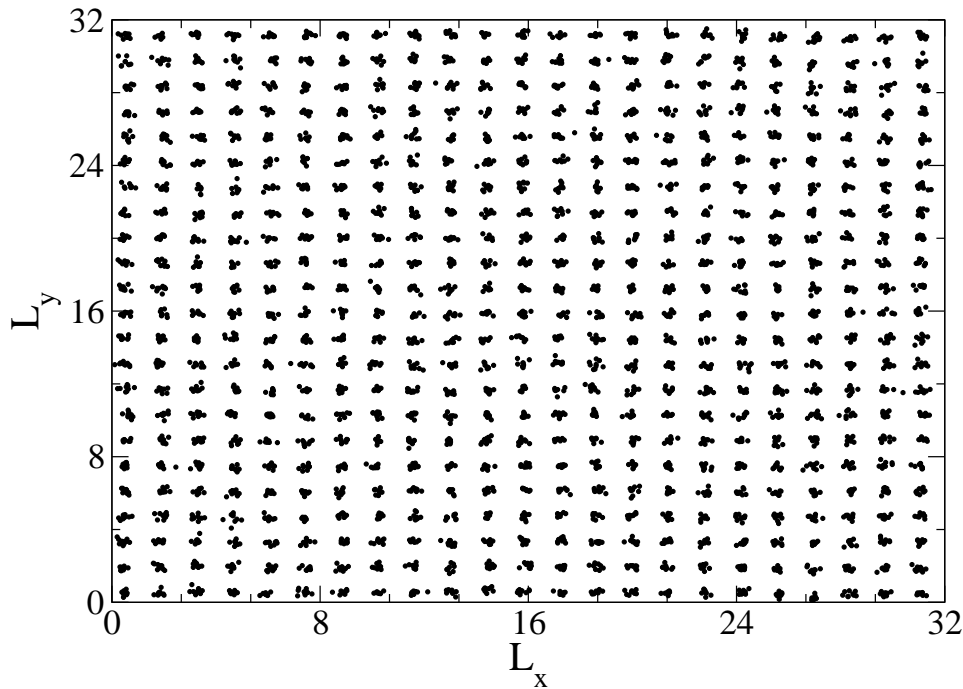}
\caption{ Ordering at small $\tau$. Initial configuration is $23 \times 23$ clouds. 
Observed configuration is $23 \times 23$. Parameters: $L/a=32$, $A=1/60$, $M\simeq4.69$, $\tau=0.0005$ and $k_BT=1.0$.}
\label{FIG_FREEZE23}
\end{center}
\end{figure}

\begin{figure}[bh!]
\begin{center}
\vspace{1cm}
\includegraphics[width=5in,angle=0]{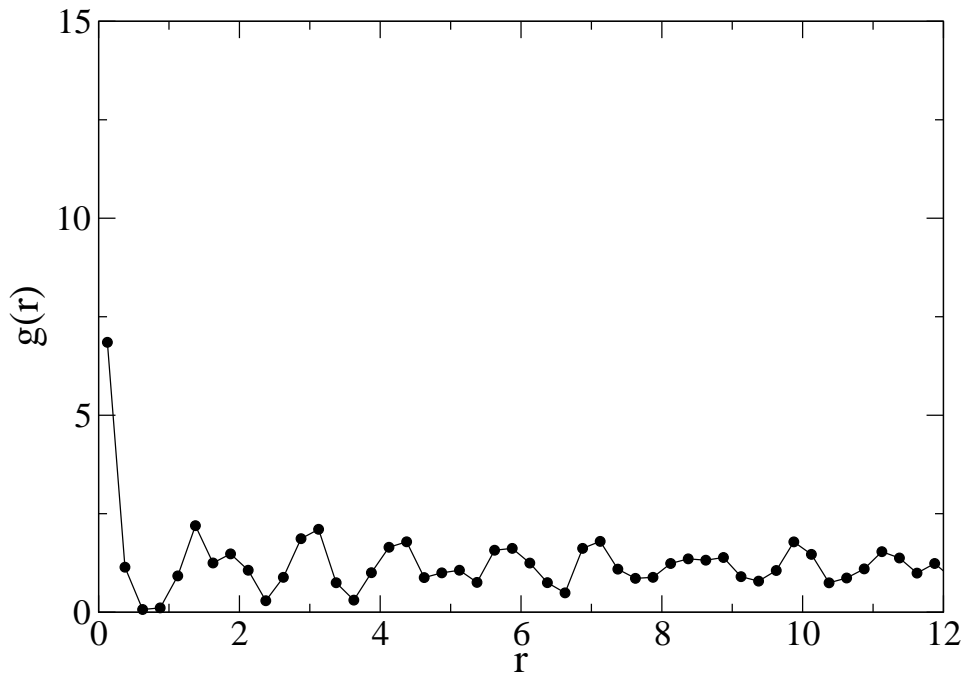}
\caption{ Pair-correlation function for the configuration shown in Fig. \ref{FIG_FREEZE23}.}
\label{FIG_PAIR23}
\end{center}
\end{figure}

\begin{figure}[th!]
\begin{center}
\includegraphics[width=5in,angle=0]{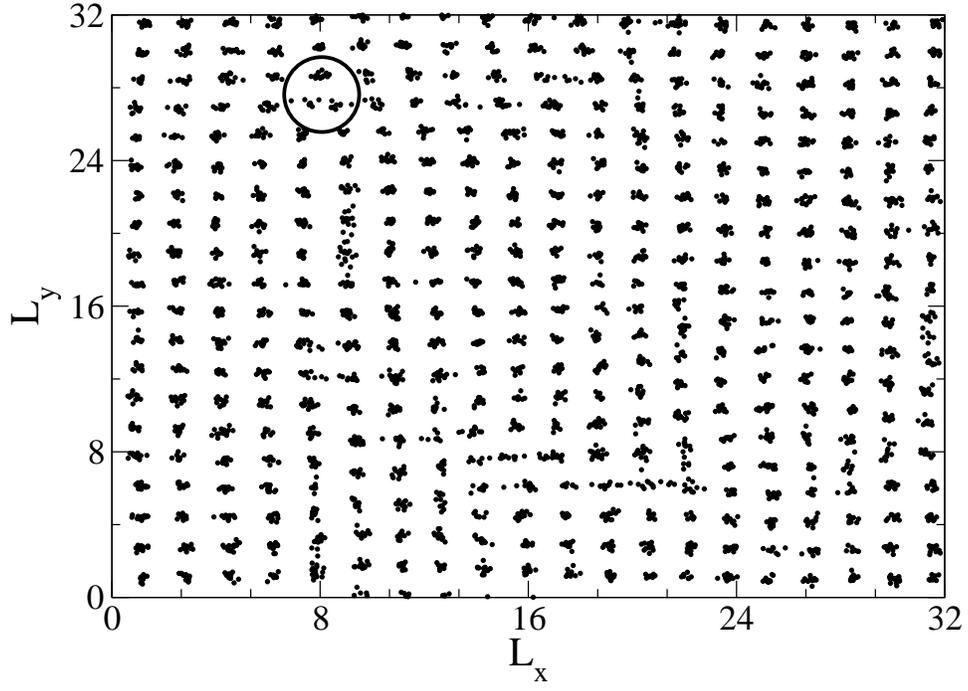}
\caption{ Ordered state with defect. Initial ordered configuration is $16 \times 16$ clouds. 
Observed configuration is $20 \times 19-20$. Parameters: $L/a=32$, $A=1/60$, $M\simeq4.69$, $\tau=0.0005$ and $k_BT=1.0$.}
\label{FIG_FREEZE20}
\end{center}
\end{figure}

\begin{figure}[bh!]
\begin{center}
\vspace{1cm}
\includegraphics[width=5in,angle=0]{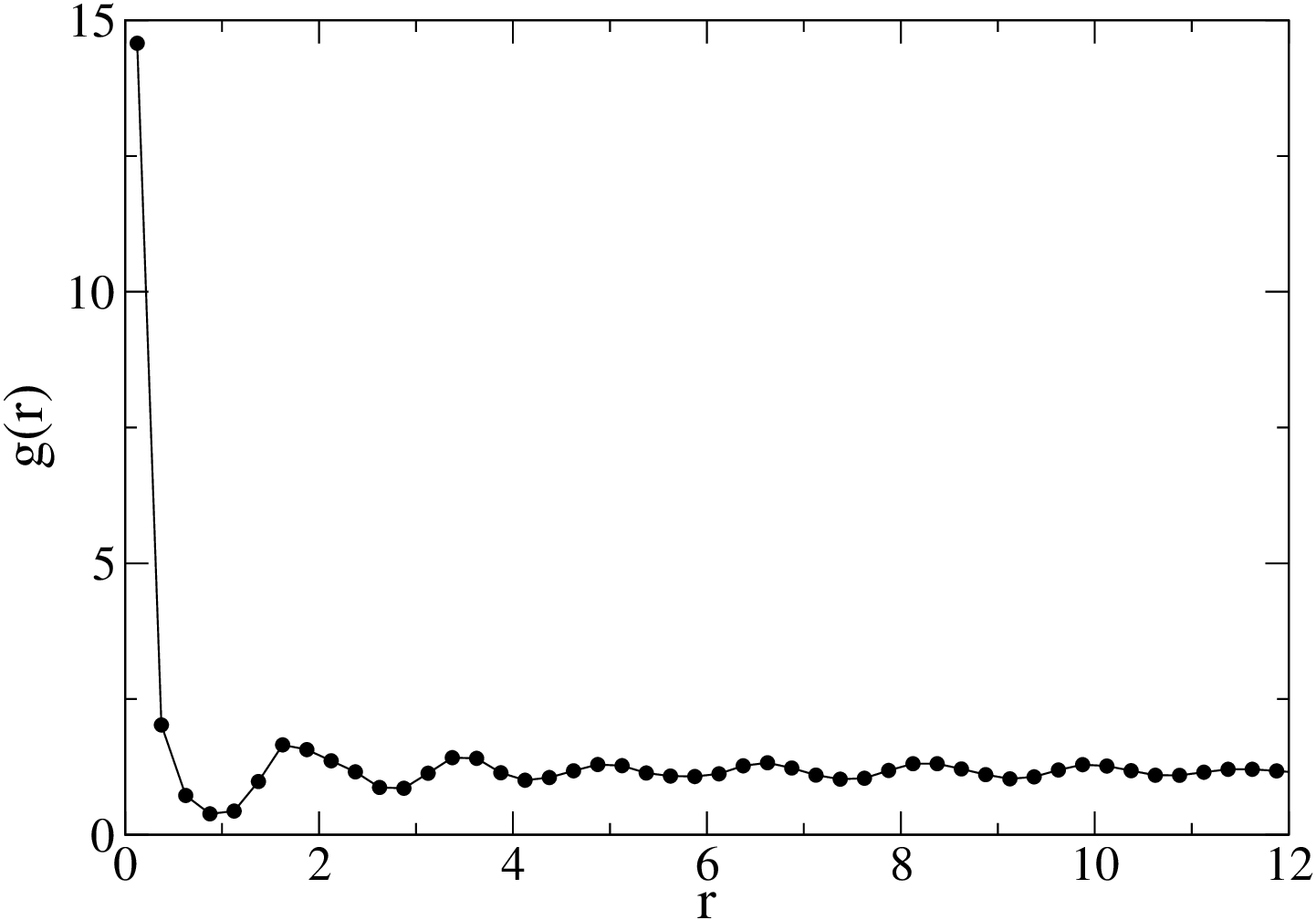}
\caption{ Pair-correlation function for the configuration shown in Fig. \ref{FIG_FREEZE20}.}
\label{FIG_PAIR20}
\end{center}
\end{figure}

\section{Caging and order/disorder transition}

When the non-ideal part of the pressure is significantly larger than the ideal part, ordering effects can be expected.
As in a real hard-sphere gas, both parts scale with temperature in the same way (for small $A$).
Hence, as in a real system which does not have an energy scale,
we can assume that changing the temperature alone will not lead to
an order/disorder transition ---  unless of course temperature becomes so large that the nonlinear terms in the pressure become relevant.
On the other hand, the two pressure terms have different dependencies on time step $\tau$ and density.
$\tau$ can be interpreted as a parameter describing the efficiency of a collision; 
lowering $\tau$ would have a similar effect as making the spheres larger in a real system, resulting in a higher collision frequency.
Therefore, we expect caging and ordering effects if either $\rho$ is increased or $\tau$ is decreased.
This is indeed the case. For $\tau< 0.0016$ and $M\approx 5$,
an ordered cubic state is observed, as shown in
Fig. \ref{FIG_FREEZE23}.
In Fig. \ref{FIG_PAIR23}, the pair-correlation for this state is plotted.
Unlike in a real system, clouds of several particles are concentrated at locations $r_{nm}=na_x+ma_y$,
where $a_x$, $a_y$ are the periodicity in x and y-direction, respectively.
The average number of particles in these clouds will be called the cloud number.
We found that there is usually at least four particles per cloud.
These ordered structures are similar to the low-temperature phase of
particles with a strong repulsion at intermediate distances, but a soft repulsion at short distances.

\begin{figure}
\begin{center}
\vspace{2cm}
\includegraphics[width=5in,angle=0]{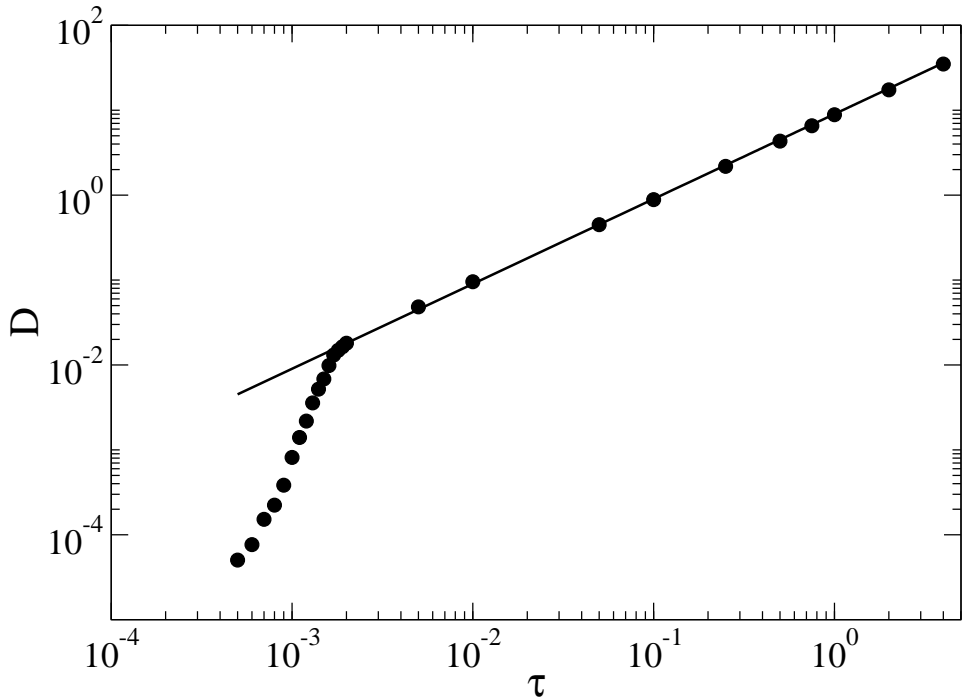}
\caption{Self-diffusion coefficient $D$ as a function of time step $\tau$. 
The solid line is the prediction of Eq. (\ref{EQ_DIFF}). Parameters: $L/a=64$, $A=1/60$, $M=5$ and $k_BT=1.0$.}
\label{FIG_DIFF}
\end{center}
\end{figure}

\begin{figure}
\begin{center}
\vspace{2cm}
\includegraphics[width=5in,angle=0]{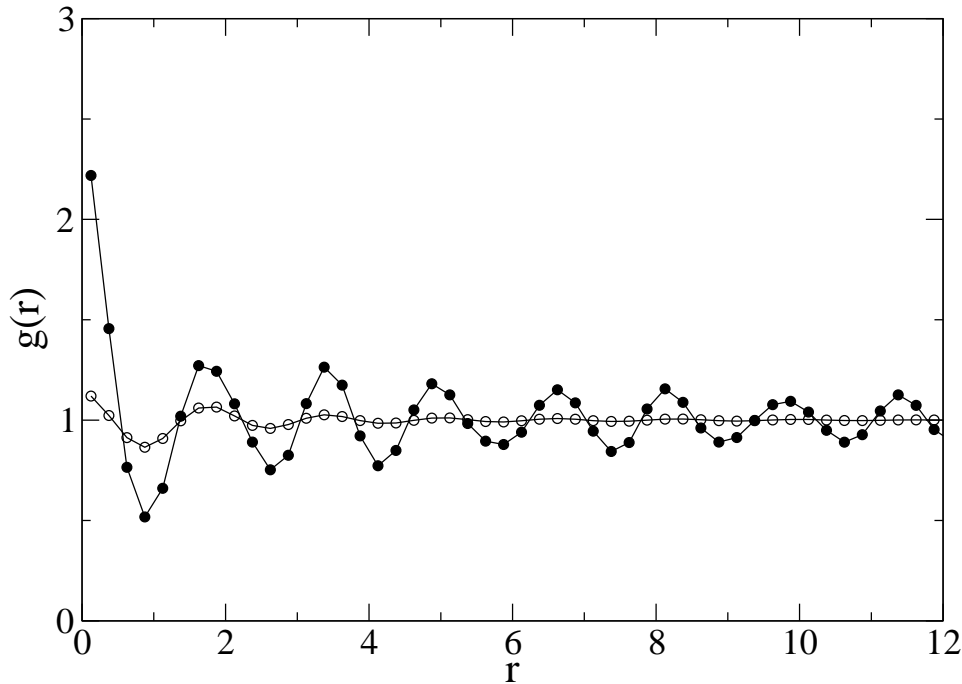}
\caption{ Pair-correlation function below and above freezing. The filled ($\smallblackcircle$) and open ($\smallwhitecircle$) circles show measurements for $\tau=0.0014$ and $\tau=0.0018$, respectively. Parameters: $L/a=32$, $A=1/60$, $M=5$ and $k_BT=1.0$.}
\label{FIG_PAIR_FREEZE}
\end{center}
\end{figure}

One of the surprising features of this crystal-like state is, that x-y symmetry can be broken, i.e. $a_x$ is not always equal to $a_y$.
Furthermore, there is the possibility of having several possible metastable crystalline states corresponding to slightly different 
lattice constants and cloud number. For instance, Fig. \ref{FIG_FREEZE23} and Fig. \ref{FIG_FREEZE20}, were
created for the same parameters but different initial conditions. While there is $23 \times 23$ clouds
in the first figure, there is $20\times 19-20$ in the other. This means also that in the latter picture
$a_x$ and $a_y$ differ slightly. Both states are relatively stable over long times. However, in Fig. \ref{FIG_FREEZE20} we see a lattice defect (circled for better visibility) and there is still activity in the lower middle part of the sample where particles seem to migrate from one cloud to another.
As expected, the lattice constants $a_x,a_y\approx 1.6$ are slightly smaller than the super cell spacing $2a$ which sets the range of the multi-particle
interaction.
In this state the diffusion coefficient becomes very small, particles are caged and can barely leave their location.
As shown in Fig. \ref{FIG_DIFF}, we measured the coefficient of self-diffusion $D$ as a function of time step $\tau$ in the long-time limit, i.e. after an eventual transition to an ordered state.
Above a critical $\tau_C\approx 0.0016$ (at $M=5$) the measured value of $D$ agrees perfectly with the theoretical prediction:
\begin{equation}
\label{EQ_DIFF}
D=k_B T\,\tau\left[
{1\over A M^{3/2}}\sqrt{\pi\over k_B T}-{1\over 2}
\right]
\end{equation}
($A$ is small and particle number fluctuations were neglected here).
However, $D$ dramatically drops below $\tau_C$ which
is exactly the value of $\tau$ below which we clearly see an ordered state.
This pronounced effect allows us to locate the transition point quantitatively.
Another way to describe the transition is by determining the pair-correlation function $g(r)$.
Fig. \ref{FIG_PAIR_FREEZE} shows that the behavior of $g(r)$ is very different slightly above and below $\tau_C$.
Below $\tau_C$ we see oscillations due to a cubic structure with long-range order with lattice constant $\approx 1.6\,a$.
To describe the transition in more detail one could plot the amplitude of the  Fourier-mode for the oscillation in $g(r)$ (not shown here).

How can we understand the stability of the ordered state? Without collisional interaction, particle clouds will broaden due to streaming, which will happen the faster the higher temperature is.
Due to the grid shift, particles at the perimeter of the clouds will more often undergo collisions with 
neighboring clouds.
These collisions provide back-scattering, particles are forced to fly back towards the center of their cloud.
Thus, a correlation between distance from cloud center and speed is built up leading to stable cloud formation.

\section{Generalization to a binary mixture}

The collision model can be easily generalized to multi-component mixtures.
Consider a binary system with two types of particles, $A$ and $B$.
In order to obtain phase separation we introduce a repulsive interaction between different kind of the particles,
but no repulsion among particles of the same kind.
This is done in the following way: Suppose a double cell is selected for a possible collision.
$A$-particles in cell 1 can now undergo a collision with $B$ particles from cell 2. Furthermore, $B$ particles from cell 1 are checked for possible collision with $A$ particles in cell 1.
The rules and probabilities for these collisions are exactly the same as in the one-component situation.
Since in a phase separated situation there is hardly any collision away from the phase boundary, additional regular 
SRD-rotations on the cell level are incorporated to mix particle momenta.

A phase separation into an $A$ and $B$ rich phase is observed above a critical prefactor $A$ of about $A_c=0.36$
for $M_A=M_B=5$, $M_{A,B}$ being the average number of $A$ and $B$ particles in a cell, respectively. 
The phase diagram does not depend on temperature as expected since similiar to a hard-sphere gas there is no
energy scale in the present model.
The 
spectrum of the interface fluctuations of droplets well above $A_c$ was analyzed, and it was found that 
they scale with wave number $k$
as $1/k^2$ as expected. 
Quantities like pressure, chemical potentials, surface tension and transport coefficients were also obtained 
analytically and 
will be published elsewhere \cite{tuzel_07}.

\section{Conclusion}
In this paper we analyzed 
a particle-based model for fluid dynamics
with effective excluded volume interactions which was
introduced in Ref. \cite{ihle_06a}.
These interactions are modeled by means of
stochastic multiparticle collisions which are biased and depend on local velocities and densities, but
exactly conserve momentum and energy locally.
The motivation for such a model and the collision rules were explained in detail.
The relaxation to thermal equilibrium was investigated analytically and numerically, and it was
shown how this relaxation was used as a guide to make the model as isotropic as possible
on a cubic grid.
A brief summary of how to calculate the pressure was already given in Ref. \cite{ihle_06a};
here we give a more detailed description of the calculation.
We also show how a microscopic formula for the pressure can be derived, which allows local
measurements of the pressure.

In addition, we outlined how a discrete-time projection operator technique developed for original SRD \cite{ihle_03} can be 
generalized to the current model, and how Green-Kubo relations can be obtained for 
the
transport coefficients. 
Numerical measurements of
the velocity and stress auto-correlation functions did
not show long-time tails even though all the simulations were done in two dimensions.
We think that this is mainly due to the large viscosity of the model, leading to amplitudes
of the tails much smaller than the numerical resolution.

In the limit of very large collision frequency, i.e. at large particle density and/or small time step $\tau$
we found an ordered cubic phase. The four-fold symmetry of this state is a numerical artifact due
to the underlying cubic grid, but the crystallization itself resembles caging
in a real gas and is a generic feature of models with soft repulsion at short distances.
This order/disorder transition was quantitatively located by measuring the self-diffusion coefficient $D$.
At the transition we saw a pronounced change of the behavior of the diffusion coefficient: $D$ 
is about one to two orders of magnitude lower in the ordered phase than it would be in a homogeneous system.
 
The ordered state consists of particle
clouds containing at least four particles. These clouds  form  metastable
cubic phases with long-range order. The lattice constant of these phases could slightly vary in x- and y-direction for the same set of parameters.
This is due to the additional degree of freedom of how many particles on average can live in a cloud.
Lattice defects were observed leading to a slow evolution from less favourable states where the lattice
constant was below or above the optimum range. This range seems to be centered around $1.6 a$.

The current model was extended to multi-component mixtures such as binary mixtures and microemulsions,
and the critical behavior was investigated.
These studies will be published elsewhere.

\section{Acknowledgement}
Support from the National Science Foundation
under Grant No. DMR-0513393 and ND EPSCoR through
NSF grant EPS-0132289 
are gratefully acknowledged.
We thank D.M. Kroll for many valuable discussions and suggestions.
We also thank A.J. Wagner for various discussions, especially about phase space contraction.

\appendix

\section{Averages of $\Delta u$ }

Averages of $\Delta u$, the mean velocity difference between two cells, show up in the calculation of the acceptance rate for a collision.
They can be measured as a test of the correct implementation of the code.
We restrict ourselves to horizontal cells, where only the x-component of velocities are needed.
In this case
$\Delta u$ is defined as
\begin{equation}
\Delta u={ v_{1,x}+v_{2,x}+...+v_{M_1, x}\over M_1 }-{v_{M_1+1,x}+v_{M_1+2,x}+..+v_{M_1+M_2,x} \over M_2} \;\;.
\end{equation}
Here $M_1$ and $M_2$ are the current particle numbers in the two involved cells, respectively.
The average over the velocity distribution at fixed $M_1$ and $M_2$ is given by
\begin{equation}
\langle |\Delta u| \rangle=\Big\langle 
\int_0^\infty\,\Delta u\,\, p_G(\Delta u)\,\,d\Delta u
\Big\rangle=\sqrt{{k_B T\gamma \over 2\pi}}
\end{equation}
where $p_G$ is given in Eq. (\ref{PRESS4}).
The particle mass is set to one as before, and $\gamma$ is defined as in Eq. (\ref{PRESS6}).

The average over the particle fluctuations, which we again assume to be Poisson-distributed, and uncorrelated on the cell level cannot be done analytically.
Let us first consider $\langle (\Delta u)^2 \rangle$ which does not have the complication of a 
square root, $\sqrt{ \gamma}$.
It involves the following averages
\begin{equation}
\label{PARTAV1}
\Big\langle\Big\langle
{1\over M_1}
\Big\rangle\Big\rangle
={\rm e}^{-M} \sum_{n=1}^\infty\,{1 \over n}\,{M^n\over n!} \;\;.
\end{equation}
The double bracket denotes the average over the particle number in a cell, while the single bracket is for averages over velocity distributions.
The r.h.s. of Eq. (\ref{PARTAV1}) can be expressed in integral form:
\begin{equation}
\label{PARTAV2}
\Big\langle\Big\langle
{1\over M_1}
\Big\rangle\Big\rangle
={\rm e}^{-M}\,\int_0^M\,{ {\rm e}^x-1\over x}\,dx=y(M)
\end{equation}
where we define a function $y(M)$.
It is assumed here that in case of $M_1=0$ the corresponding term involving this cell does not occur at all in 
$\Delta u$, for example if $M_1=0$ and $M_2=2$, $\Delta u$ would be $\Delta u=(v_{1x}+v_{2x})/M_2$ i.e. involves only 
velocities from cell 2. In other words, we assume $1/M_1$ to be zero for $M_1=0$.

From Eq. (\ref{PARTAV2}) it follows that $y(M)$ satisfies the following linear differential equation
\begin{equation}
\label{PARTAV3}
{dy\over dM}+y={1-{\rm e}^{-M} \over M}
\end{equation}
For $M\ll 1$ we obtain the following approximative solution:
\begin{equation}
\label{PARTAV4}
y(M)\sim M-{3\over 4} M^2+{11\over 36} M^3-...
\end{equation}
whereas for large $M$, $M\gg 1$ the asymptotic solution is
\begin{equation}
\label{PARTAV5}
y(M)\sim -{\rm e}^{-M}\,{\rm ln}M+{1 \over M}+{1\over M^2}+{2\over M^3}+{6 \over M^4}+...
\end{equation}

Let us now consider 
$\langle |\Delta u| \rangle$ 
where the average over the particle numbers cannot be done independently for every cell but
involves a double sum:
\begin{equation}
\label{PARTAV6}
\Big\langle\Big\langle
\sqrt{{1\over M_1}+{1\over M_2}}
\Big\rangle\Big\rangle
={\rm e}^{-2M} \sum_{n,m=1}^{\infty}\,\sqrt{{1\over n}+{1\over m}} \,{M^{n+m}\over n!m!}
+2{\rm e}^{-2M} \sum_{n=1}^{\infty}\,\sqrt{{1\over n}} \,{M^n \over n!}
=g(M)
\end{equation}
where the second sum comes from the special treatment if one of the two cells is empty.

For very large $M$, $M_1$ and $M_2$ are simply replaced by their average value $M$
and one obtains
\begin{equation}
\label{PARTAV7}
g(M)\sim \sqrt{{2\over M}}
\end{equation}

For small $M$ just the first few terms in the sums are taken into account and one finds for $M\ll 1$:
\begin{equation}
\label{PARTAV8}
g(M)\sim 2M-(4-2\sqrt{2})\,M^2+(8+\sqrt{22/3}-4\sqrt{2})\,M^3+...
\end{equation}
Hence, the behavior of  
$\langle |\Delta u| \rangle /\sqrt{k_B T}$ is propotional to 
$M \sqrt{2/\pi} -M^2(4-2\sqrt{2})/\sqrt{2 \pi}$ at small $M$, reaches a maximum around $M=1$ and behaves as $\sim 
1/\sqrt{\pi M}$ at very large $M$.
This non-monotonic behavior suggests that functions of $\gamma$ 
should not occur in the equation of state but should be cancelled by an appropriate collision probability.
Without the cancellation there might be unphysical phase transitions at densities around $M=1$.

\end{document}